\documentclass[sigconf,nonacm]{acmart}

\usepackage{amsmath,amsthm}

\usepackage{amssymb}
\usepackage{array}
\usepackage{booktabs}
\usepackage{graphicx}
\usepackage{microtype}
\usepackage{enumitem}
\usepackage{xspace}
\usepackage{cleveref}
\usepackage{algorithm}
\usepackage{algpseudocode}
\usepackage{tikz}
\usetikzlibrary{arrows.meta,positioning,fit,backgrounds}

\newtheorem{definition}{Definition}
\newtheorem{theorem}{Theorem}
\newtheorem{corollary}{Corollary}
\newtheorem{lemma}{Lemma}
\newtheorem{proposition}{Proposition}
\newtheorem{remark}{Remark}
\crefname{definition}{Definition}{Definitions}
\crefname{theorem}{Theorem}{Theorems}
\crefname{corollary}{Corollary}{Corollaries}
\crefname{proposition}{Proposition}{Propositions}
\crefname{lemma}{Lemma}{Lemmas}
\crefname{remark}{Remark}{Remarks}

\newcommand{\budget}{B}                   %
\newcommand{\totalviews}{V}               %
\newcommand{\itemidx}{s}                  %
\newcommand{\creator}{c}                  %
\newcommand{\sensitivity}{\delta}         %

\newcommand{\repro}{\mathcal{R}}          %
\newcommand{\Vorg}{V_{\mathrm{org}}}      %
\newcommand{\subs}{\Lambda}               %
\newcommand{\corpus}{K}                   %
\newcommand{\turnover}{\omega}            %
\newcommand{\creatorresp}{g}              %
\newcommand{\reprooff}{\repro_{\mathrm{off}}}   %
\newcommand{\qoff}{q_{\mathrm{off}}}            %
\newcommand{\subsoff}{\subs^{*}_{\mathrm{off}}} %
\newcommand{\corpusoff}{\corpus^{*}_{\mathrm{off}}} %

\newcommand{\bpara}[1]{\par\vspace{1pt}\noindent\textbf{#1}\hspace{0.5em plus 0.1em minus 0.05em}}

\newif\ifextendedbuild
\extendedbuildfalse

\extendedbuildtrue

\newcommand{\appref}[1]{Appendix~\ref{#1}}
\newcommand{\resref}[2]{#1~\ref{#2}}
\newcommand{\suppname}{appendix}

\begin{document}
\raggedbottom

\title{Content Exploration Beyond the Feed: Creator Supply and the Shared Corpus}
\author{Yuanyuan Shen}
\affiliation{\institution{Snap Inc.}\city{New York}\state{NY}\country{USA}}
\email{yshen2@snapchat.com}

\author{Yiren Yan}
\affiliation{\institution{Snap Inc.}\city{Palo Alto}\state{CA}\country{USA}}
\email{yyan5@snapchat.com}

\author{Wenjie Li}
\authornote{Work done while at Snap Inc.}
\affiliation{\institution{Snap Inc.}\city{Palo Alto}\state{CA}\country{USA}}
\email{wenjie.li@snapchat.com}

\author{Chunhui Zhu}
\affiliation{\institution{Snap Inc.}\city{Palo Alto}\state{CA}\country{USA}}
\email{chunhui.zhu@snapchat.com}

\begin{abstract}
Industrial recommenders give new content initial views through budgeted exploration, then use early performance to decide further delivery. On many short-video platforms, exploration is the primary route by which new videos reach viewers. Viewer-side tests measure consumption, while the published budget objectives we review omit creator response. We analyze four experiments on a major short-video platform. An eight-month creator ablation finds that production exploration raises videos posted per creator by 8.55\% and creators posting at least once by 7.10\% relative to a minimal floor. A budget-matched reallocation raises creator participation with no detectable short-run viewer-side change. A year-long viewer ablation separately finds 1.74\% more video views but 2.13\% less view time.
A delivered view creates immediate feed value, can trigger organic take-up, and can induce creator supply. Take-up and supply replenish a shared corpus, creating two measurement limits. Viewer-side A/B tests cancel the corpus effect when both arms consume the same corpus. Giving each arm its own corpus avoids cancellation, but turnover still controls the horizon. If the corpus turns over at rate $\turnover$ per posting cycle, a $t$-cycle experiment expresses at most $\turnover t$ of the eventual corpus effect. More users reduce noise without making the corpus turn faster. Before the corpus path visibly bends, data cannot distinguish a modest fast effect from an arbitrarily large slow one, so a valid confidence interval may lack a finite upper endpoint.
As predicted, the three-week co-diverted experiment cannot determine whether the eventual corpus effect is positive or negative. Within the window, it identifies the direct feed effect, and an exploratory cohort analysis detects organic lift after exploration ends. Together, the experiments establish a positive creator response, measure the gross corpus flow visible within three weeks, and show the design and duration needed to identify total value.

\end{abstract}



\maketitle

\section{Introduction}
\label{sec:intro}

Short-video platforms receive enormous volumes of new uploads, most of which
initially reach few viewers. Ranking models favor content with an observed
response, so new videos lack the history needed to compete. \emph{Exploration}
gives them initial views. On many short-video platforms, exploration is the
primary distribution channel for most new videos.

\begingroup
\setlength{\emergencystretch}{2em}
Platforms commonly give each new video an \emph{exposure budget} and use early
performance to decide whether to release more views~\citep{shen2025aliboost,gudmundsson2026pinequalizer,wang2025item,jeon2025epinet,chen2025kuaishou}.
We study systems that allocate exposure in stages before each video's
exploration period ends. A video
that clears an early bar receives more exposure. A video that misses it leaves
the exploration pool (\S\ref{sec:related}).
\par
\endgroup

This design creates an evaluation problem. Viewer-side A/B tests can show
mixed or negative consumption effects even though exploration supplies most
new-video distribution. Related systems report the same short-window
pattern~\citep{su2024exploration,chen2021values}. Longer one-sided tests share
the same corpus across arms, so its contribution cancels at every horizon.

On the major short-video platform we study, an eight-month creator ablation
shows that production exploration raises videos posted per creator by 8.55\%
and creators posting at least once by 7.10\% relative to a minimal floor.
A separate budget-matched reallocation raises creator participation with no
detectable short-run viewer-side change. A year-long viewer ablation finds
1.74\% more video views and 2.13\% less view time relative to disablement. The
viewer path shows no sustained buildup over twelve months. These results
establish a positive creator response and isolate the immediate viewer trade-off.

Prior isolated-submarket experiments establish two corpus links:
exploration expands the discoverable corpus, and random thinning shows that
reducing the corpus lowers satisfied-user counts~\citep{su2024exploration}.
We ask how a deployed budgeted mechanism can measure the resulting value. The
creator ablation measures the posting response that replenishes the corpus,
and the viewer ablation measures the immediate feed trade-off. Co-diversion
jointly randomizes matched creator and viewer submarkets, so each arm develops
its own corpus. Our three-week probe lands at the estimation floor predicted by
the theory. At that horizon, a cohort analysis measures gross corpus flow,
while the aggregate asymptote remains unsigned.
\Cref{fig:causal-map} maps the three channels to the designs that identify
them.
\ifextendedbuild\else\newpage\fi

\newcommand{\causalmapbody}{%
\begin{tikzpicture}[
  >={Latex[length=2.4mm,width=2.1mm]},
  box/.style={draw, rounded corners=2.5pt, align=center, minimum height=7.5mm,
              inner xsep=5pt, font=\small, line width=0.6pt},
  src/.style={box, fill=black!12, draw=black!75, line width=0.8pt},
  fastbox/.style={box, fill=orange!16, draw=orange!55!black},
  slowbox/.style={box, fill=blue!8, draw=blue!45!black},
  stockbox/.style={box, fill=blue!15, draw=blue!45!black, line width=0.8pt},
  abtag/.style={draw=black!60, dashed, rounded corners=1.5pt, fill=white,
                font=\scriptsize, inner sep=2.6pt, align=center},
  flow/.style={->, line width=0.75pt, draw=black!80},
  measure/.style={densely dotted, line width=0.5pt, draw=black!60},
  elab/.style={font=\scriptsize, inner sep=1.2pt, fill=white}
]
\node[src]      (explore) at (0,1.6)   {exploration\\views};
\node[fastbox]  (feed)    at (5.2,1.6) {current feed\\value};
\node[slowbox]  (creator) at (0,0)     {creator\\posting};
\node[stockbox] (corpus)  at (2.6,0)   {shared\\corpus};
\node[slowbox]  (future)  at (5.2,0)   {future viewer\\value};

\begin{scope}[on background layer]
\node[fill=blue!4, draw=blue!20, rounded corners=4pt,
      inner xsep=7pt, inner ysep=6pt, fit=(creator)(corpus)(future)]
      (slowband) {};
\end{scope}
\node[font=\scriptsize\itshape, text=blue!35!black, anchor=south]
      at ([xshift=0.7cm]slowband.north) {slow channels};

\draw[flow] (explore) -- node[elab, above] {immediate} (feed);
\draw[flow] (explore) -- node[elab, left]  {response} (creator);
\draw[flow] (explore) -- node[elab, above, sloped, pos=0.45] {take-up} (corpus);
\draw[flow] (creator) -- node[elab, above] {new posts} (corpus);
\draw[flow] (corpus)  -- node[elab, above] {serves} (future);

\node[abtag] (vtag) at (5.2,0.82)  {viewer-side A/B};
\node[abtag] (ctag) at (0,-1.05)   {creator-side A/B};
\node[abtag] (dtag) at (5.2,-1.05) {co-diverted A/B};
\draw[measure] (vtag) -- (feed);
\draw[measure] (ctag) -- (creator);
\draw[measure] (dtag) -- (future);
\end{tikzpicture}%
}

\newcommand{\causalmapcaption}{Value channels and identifying designs. Exploration has an immediate feed effect,
can change creator supply, and feeds a shared corpus through both take-up and
new posts. The shaded channels express slowly. Each dashed design reaches a
different channel. A one-sided design shares the corpus across arms, so that
channel cancels. A separate budget-matched co-diverted variant perturbs allocation at
fixed nominal budget and measures the creator response (\S\ref{sec:reallocation}).}

\ifextendedbuild
\begin{figure*}[t]
\centering
\resizebox{0.68\textwidth}{!}{\causalmapbody}
\caption{\causalmapcaption}
\label{fig:causal-map}
\end{figure*}
\else
\begin{figure}[t]
\centering
\resizebox{\columnwidth}{!}{\causalmapbody}
\caption{\causalmapcaption}
\label{fig:causal-map}
\end{figure}
\fi

Published cold-start systems score and budget videos with viewer and corpus
quantities such as predicted discoverability and click-through. None of the
systems we review includes the creator's later posting response in its
\emph{budget} objective. Production feed ranking outside this class has modeled
creator incentives for years~\citep{tu2019feedback}
(\S\ref{sec:related}).

\begingroup
\setlength{\emergencystretch}{2em}
Cold-start systems must decide within the delivery window whether a video
receives more exposure or leaves the pool. At that point, the system observes
only fast viewer signals. The delayed posting response and corpus turnover
therefore require offline estimates. Historical creator experiments and
corpus-age data supply those estimates, and budget objectives can use them as
priors. A horizon-matched creator
experiment identifies the supply response. Co-diversion retains the
shared-corpus channel (\S\ref{sec:creator-ablation} and \S\ref{sec:probe}).
\par
\endgroup

We model the ecosystem that exploration feeds and test its channels with four
experiments on a major short-video platform (\S\ref{sec:budget-problem}). The
linear model supports the value accounting. The timing results require only
corpus turnover and survive without linearity (\S\ref{sec:measurement}).
The two long-running ablations compare the mechanism with disablement or a
minimal floor, so their estimands are invariant to allocator details. The
budget-matched experiment separately changes allocation at a fixed nominal
budget. We contribute a valuation and measurement framework for these rules.
Algorithm design lies outside our scope. We make three contributions:
\begin{enumerate}[leftmargin=*, nosep]
  \item \textbf{Signed creator response.} An eight-month ablation of production
  exploration raises videos posted per creator by 8.55\% and creators posting
  at least once by 7.10\% above a minimal floor. The published cold-start budget
  objectives we review omit this response.
  A budget-matched reallocation raises creator participation with no detectable
  short-run viewer cost (\S\ref{sec:experiments}).
  \item \textbf{Value and identification.} We separate a delivered view's value
  into the immediate view, organic take-up, and induced creator supply, then map
  each channel to its identifying randomization. A posting loop with gain
  $\repro$ amplifies posting by $1/(1-\repro)$. One-sided designs identify the
  direct and supply channels but
  exclude the shared corpus by construction. These are distinct estimands, and
  the designs determine which net signs are identified
  (\S\ref{sec:dynamics}, \S\ref{sec:value}, \S\ref{sec:experiments}).
  \item \textbf{Corpus horizon.} Only isolated cells retain the corpus
  contribution in their contrast, and they must wait for the corpus itself to
  turn over. Under a monotone stock transition with turnover $\turnover$, a
  $t$-cycle experiment expresses at most $\turnover t$ of
  the asymptote. Additional users reduce noise but cannot advance that
  transition. The three-week probe lands at the predicted estimation floor. An
  exploratory cohort analysis measures gross corpus flow at the feasible
  horizon, while the aggregate asymptote and net value after displacement
  remain unresolved. This result separates what three weeks can identify from
  what requires a longer isolated trajectory
  (\Cref{thm:horizon,prop:detection}; \S\ref{sec:probe}).
\end{enumerate}

The long-running ablations estimate a viewer-feed effect after substitution
and an aggregate posting response. These effects use
different units, so we report them separately. A total-value estimate requires
a common scale. Unlike
one-off traffic grants, the ablations measure a standing mechanism on both market
sides over months. The shared-corpus value requires co-diversion at the
prescribed horizon~\citep{su2024exploration}. Below a curvature threshold, that
design can lower-bound an expressed increment while the asymptote still lacks a
finite upper bound (\S\ref{sec:probe}).

\section{Related Work}
\label{sec:related}

\bpara{Content Cold-Start in Recommender Systems.}
Cold-start is among collaborative filtering's oldest
problems~\citep{schein2002methods,lops2011content,lee2019melu}, met on video
platforms with a dedicated exploration phase before full
recommendation~\citep{covington2016deep}. We treat that phase as a first-class
distribution channel carrying creator as well as viewer value.

\bpara{Long-Term and Slow-Expressing Effects.}
\begingroup
\setlength{\emergencystretch}{2em}
Short experiments are known to mismeasure long-run impact, with slow
expression treated as a \emph{user-side} phenomenon to be estimated around.
Existing methods include cohort and stepped-wedge designs for user
learning~\citep{hohnhold2015longterm}, surrogate
indices~\citep{athey2019surrogate}, and temporal designs for
carryover~\citep{basse2023longterm}. Here the slowness is structural, a
two-sided supply loop with a corpus stock, and the question is when the
long-run effect is observable at all. Together, \Cref{thm:horizon,cor:duality},
the $N^{-1/3}$ detection law of \Cref{prop:detection}, and the $N^{-1/5}$
estimation law of \Cref{thm:curvature} bound what a finite-length experiment can see of the equilibrium effect. The gain--relaxation
relation is classical in Markov mixing and critical slowing down. The
phenomenon of unbounded confidence sets behind \Cref{thm:curvature} is classical
in weakly identified models~\citep{gleser1987nonexistence,dufour1997impossibility}.
We specialize these results to the exponential-approach family, derive an
operational curvature constant and horizon laws, and tie them to exploration's
equilibrium value.
\par
\endgroup

\bpara{Industrial Cold-Start Allocation.}
Published cold-start systems share two features. New items receive a
per-item exposure allocation, and continued exposure is gated on early
performance. \citet{wang2025item} allocate per-item traffic under a global
budget with a floor and ceiling by inverting a learned discoverability
probability. \citet{wang2025itemcentric} screen candidates using a Beta
posterior that balances satisfaction against item quality.
\citet{jeon2025epinet} reserve quota below an impression threshold, and
\citet{chen2025kuaishou} pair an exposure floor with flow steered toward
quality content. The full mechanism analyzed here adds a bounded delivery
window with escalation and withdrawal, as in the tiered system of
\citet{shen2025aliboost}. The platform we study runs this mechanism class,
although published accounts differ in how much of its structure they report.

\bpara{Objectives and Their Timing.}
The systems differ in what determines the budget.
\citet{wang2025item} maximize the number of items becoming ``discoverable
post exploration.'' The tiered system above budgets on predicted
click-through and gates escalation on realized CTR. None of the systems
reviewed above includes a term for
the creator's subsequent supply behavior \emph{in its budget objective}, a
statement about the published record that the structural argument of
\S\ref{sec:intro} predicts. That system motivates its design partly by
producer incentives, yet no producer variable enters its budget rule.

The gap is specific to this class. \citet{tu2019feedback} reshape a production
feed's feedback distribution to nurture creators, raising creation at no
measured consumer cost. \citet{wang2025dynamic} derive allocation policies
that price creators' future growth. The settings differ in cadence. Feed
ranking can adapt against long-run outcomes, whereas a cold-start budget must
be set, escalated, and withdrawn within a bounded window whose exit gate
returns to viewer-side evidence. Removing that gate reportedly degrades
viewer-side metrics~\citep{shen2025aliboost}, yet its creator-side cost is
unmeasured.

Two prior results sit closest to ours.
With two separate interventions, \citet{su2024exploration} establish that
exploration grows the discoverable corpus through joint user and corpus
co-diversion. They then show that random corpus thinning costs satisfied
users, with the benefit accruing over months. \citet{shen2026isolation}
characterize the cost imposed by co-diversion itself. We
identify the creator-supply response to a standing budgeted mechanism,
separate structural cancellation from horizon attenuation, and derive the
design and information limits that govern direct measurement of the
corpus-mediated consequence.

\bpara{Adjacent Lines.}
Prior systems in this family used bandit exploration~\citep{li2010contextual,chen2021values}.
It valued a view for information about an arm. Budgeted allocation instead
prices the view as a distribution quantity (\Cref{prop:gradient}). Ad delivery under
budgets~\citep{mehta2007adwords,balseiro2019learning} contributes the
mechanics but prices impressions in a common currency and omits the creator's
behavioral response. Field experiments on two-sided platforms
establish that exposure causally shapes creator supply.
Randomized traffic raises production by $5.87\%$~\citep{hu2024traffic}.
Additional impressions produce higher-quality and more diverse output~\citep{xia2025supporting},
as do peer awards~\citep{burtch2022peer}. Theory connects demand allocation to
participation~\citep{rochet2003platform,caillaud2003chicken,bhargava2022creator,qian2024digital}.
Closest to our reading, \citet{yao2024unveiling} formalize the trade-off in
theory and simulation. Our ablations price a standing production mechanism
on both market sides over months. Prior experiments study one-off traffic
grants over weeks. Fairness work on provider-side
exposure~\citep{abdollahpouri2019managing,burke2017multisided} studies
minimum-exposure guarantees as commitments, while this class grants floors
for operational reasons.

\section{The Mechanism and Its Ecosystem}
\label{sec:problem}

\subsection{Budgeted Exploration as Practiced}
\label{sec:budget-problem}

Creators submit content items into an \emph{exploration pool} $\mathcal{P}_t$.
Each item $\itemidx$ remains eligible in the pool for at most $T_{\max}$, its
time-to-live (TTL). The pool competes for an aggregate exploration capacity
of $\totalviews_t$ views. A
\emph{view budget} $\budget(\itemidx) \in \mathbb{Z}_{\geq 0}$ is the target
number of exploration views to deliver to $\itemidx$ within its TTL, subject to
capacity and continued eligibility.

\begin{definition}[Budgeted Exploration Problem]
Given pool $\mathcal{P}_t$, capacity $\totalviews_t$, and a value function
$U(\itemidx, \budget)$ for delivering $\budget$ views to content item $\itemidx$, choose
budgets to
\begin{equation}
\begin{aligned}
\max_{\{\budget(\itemidx)\}_{\itemidx \in \mathcal{P}_t}} \;\; & \sum_{\itemidx \in \mathcal{P}_t} U(\itemidx, \budget(\itemidx)) \\[-1pt]
\text{s.t.} \;\; & \sum_{\itemidx \in \mathcal{P}_t} \budget(\itemidx) \leq \totalviews_t,
\quad B_{\min} \leq \budget(\itemidx) \leq B_{\max} \;\; \forall \itemidx,
\end{aligned}
\label{eq:budgeted}
\end{equation}
where $B_{\min}$ is a minimum budget guarantee (cold-start views) for every
content item and $B_{\max}$ a per-item ceiling.
\end{definition}

Industrial practice resolves this generically
(\S\ref{sec:related}). Every eligible item receives the
guarantee $B_{\min}$ before any viewer response exists. Views above it,
\emph{escalation}, are conditioned on early response inside the window.
Delivery of the remainder stops, \emph{withdrawal}, when the budget is
exhausted, expires, or repeatedly misses viewer-response criteria.
Both decisions must rely on the fastest-expressing
quantities available, and \Cref{thm:horizon} will show that decisions made
on those quantities are exactly the ones blind to the creator-supply and
corpus-stock channels of \Cref{prop:gradient}. The platform we measure runs
a standard member of this class. How budgets are set is platform-specific.
The theory conditions on the resulting budgets, while the ablations compare
the mechanism with its absence or a minimal floor. Both analyses apply without
specifying the platform's particular budget-setting rule.

Budget assignment is decoupled from delivery. The contract requires only that
each content item receive its target views from relevant viewers within its TTL,
and a dedicated funnel, ranking boosts, quotas, or guaranteed-delivery
pacing honor it equally, so the valuation and evaluation transfer to any of
them.

\subsection{An Ecosystem Model}
\label{sec:dynamics}

Whether budgeted exploration pays for itself is put to viewer-side A/B tests,
whose consumption results can be mixed or negative
(\S\ref{sec:viewer-ablation}). The viewer-side contrast identifies only the
immediate feed trade-off.
Feedback value compounds over periods and grows with persistence. No short
window can observe a slow effect, whether it is an A/B test's measurement
window or the mechanism's own gate.

\bpara{States.}
We work in discrete time. One \emph{period} is a submission-feedback cycle
that includes submission, exploration views within TTL, and the creator's next
posting decisions. Its calendar length is a platform property external to the model.
Two state variables describe the ecosystem at period $t$. The first is $\subs_t$, the new
submission rate or \emph{supply}. The second is $\corpus_t$, the \emph{organic corpus} of
content items remaining in organic distribution after their exploration window.

\bpara{Model ingredients.}
Six ingredients parameterize the model. The baseline submission rate is
$\subs_0$. The \emph{creator response} $\creatorresp(\cdot)$ gives expected
next-period submissions as a function of views received. We assume it is
increasing and \emph{concave}, consistent with the saturating responses
reported in field experiments~\citep{hu2024traffic,xia2025supporting}.
Expected total views are $\bar{V}(\budget) = \budget + q(\budget)\,\Vorg$.
This expression includes $\budget$ delivered exploration views plus, with \emph{organic take-up}
probability $q(\budget)$, a further $\Vorg$ organic views. We assume $q$ is
nondecreasing and all view-value weights are nonnegative. The corpus turns
over at per-period rate $\turnover \in (0,1)$. Viewer value is $\eta$ per
corpus unit and period, so aggregate consumption (total view time, say) is
locally affine in the stock, $E_t = \eta\,\corpus_t + \text{const}$, the
constant carrying consumption outside the corpus.
Writing $\eta_{\mathrm{o}}$ for gross value per organic view, our
normalization sets $\eta/\turnover=\eta_{\mathrm{o}}\Vorg$: both sides are
the lifetime organic value of one taken-up item.

\bpara{Dynamics.}
These ingredients give a pair of linear recursions.
\begin{equation}
\subs_{t+1} = \subs_0 + \repro\,\subs_t,
\qquad
\corpus_{t+1} = (1-\turnover)\,\corpus_t + q(\budget)\,\subs_t .
\label{eq:dynamics}
\end{equation}
The first says next period's supply is the baseline plus a fed-back share of
this period's supply. The second says the corpus retains a fraction
$1-\turnover$ and gains the taken-up share of new submissions.

\ifextendedbuild
Where the recursions settle, and what one more delivered view is worth once
the loop has run its course, follow directly.

\begin{proposition}[Exploration multiplier]
\label{prop:multiplier}
For $\repro < 1$, the dynamics~\eqref{eq:dynamics} converge to
\[
\subs^* = \frac{\subs_0}{1-\repro},
\qquad
\corpus^* = \frac{q(\budget)}{\turnover}\,\subs^*,
\]
and one additional delivered view yields $\sensitivity/(1-\repro)$
cumulative additional submissions over the ecosystem's response. Every induced
submission attracts views that induce further submissions, producing the multiplier.
\end{proposition}
\begin{proof}
The affine recursion for $\subs_t$ contracts to its unique fixed point
$\subs_0/(1-\repro)$. Substituting into the $\corpus$ recursion gives
$\corpus^*$. A marginal view induces $\sensitivity$ submissions, which
induce $\sensitivity\repro$, then $\sensitivity\repro^2$, summing to
$\sensitivity/(1-\repro)$.
\end{proof}

\fi

\bpara{Gain and sensitivity.}
The creator
response enters through two distinct quantities. The \emph{loop gain} is
$\repro \triangleq \creatorresp\bigl(\bar V(\budget)\bigr)$, the expected
number of \emph{further} submissions one submission induces and the
eigenvalue of the supply recursion. The \emph{sensitivity} is
$\sensitivity \triangleq \creatorresp'(\bar V(\budget))$. It is the marginal
posting response to one additional delivered view, and the quantity that
values a view. The recursions are mean-field approximations. Both $\sensitivity$ and $q$ are population
averages, and the subscripted $\sensitivity_\creator$, $q'_\itemidx$ of
\Cref{prop:gradient} are their individual-level counterparts. A healthy
platform is \emph{subcritical}, with $\repro < 1$. Each submission then induces less
than one additional submission, so the supply loop remains stable. The budget $\budget$ is the
lever through which the platform sets $\repro$.

\bpara{Supply loop and corpus stock.}
Stacked as $x_t = (\subs_t, \corpus_t)$, the recursions read
$x_{t+1} = M x_t + u$ with eigenvalues $\repro$ (the \emph{supply loop}) and
$1-\turnover$ (the \emph{corpus stock}). The two play different roles. The
supply loop governs how much a view is worth. The corpus stock governs how
long an experiment must run before that value is visible. No result below
requires the corpus to be slower than the supply loop: because
$1-\turnover$ is mechanical turnover and involves $\creatorresp$ nowhere,
the horizon bound holds for any creator response and any ordering of the
two eigenvalues (\Cref{thm:horizon}).

\begin{remark}[Fixed capacity]
\label{rem:capacity}
Fixing the per-item budget is compatible with finite capacity. Under
\eqref{eq:budgeted}, a larger pool lowers views per item through
$\totalviews/\subs$, adding negative feedback. Ignoring that congestion can
overstate the supply multiplier. The timing statements remain unchanged because
they depend on $\turnover$ alone. The local elasticity derivation
is in \appref{app:dynamics}. Concavity is needed only near current budgets,
where the field evidence above is most direct.
\end{remark}

\bpara{The value of a view and its multiplier.}
The dynamics converge to $\subs^* = \subs_0/(1-\repro)$ and
$\corpus^* = q(\budget)\,\subs^*/\turnover$. One marginal delivered view
induces $\sensitivity$ submissions, whose descendants form the geometric sum
$\sensitivity\sum_{k\geq0}\repro^k=\sensitivity/(1-\repro)$
(\resref{Proposition}{prop:multiplier}). Valuing a view by its direct
response alone therefore understates its worth by the factor
$1/(1-\repro)$, which grows without bound as the platform approaches
criticality, $\repro \to 1$.

\begin{remark}[Scope of the linear model]
\label{rem:scope}
\eqref{eq:dynamics} is the tangent description of a subcritical ecosystem
at steady state, including $\Vorg$. The \emph{multiplier} applies locally at
that steady state. \Cref{thm:horizon}'s bound carries no such restriction. The
$\eta$'s are gross values. An on/off
contrast measures the gap \emph{net} of displaced impressions, the
decision-relevant quantity. Additional nonnegative feedbacks would raise the
spectral radius of $M$. Opposite-sign feedbacks such as crowd-out or quality
dilution can lower it and lie outside this model
(\appref{app:dynamics}).
\end{remark}

\section{Value and Measurability}
\label{sec:theory}

\subsection{What a Delivered View Is Worth}
\label{sec:value}

The multiplier of \S\ref{sec:dynamics} values the supply loop as a whole. An
allocator needs something finer, the value of one more view to one particular
content item. Under the same model, that value separates into three channels. They are
the posting that the view induces, the view itself, and organic take-up.

\begin{proposition}[Marginal value of a view]
\label{prop:gradient}
Under~\eqref{eq:dynamics}, the cumulative consumption value of one
additional view delivered to content item $\itemidx$ from creator $\creator$ with
budget $\budget(\itemidx)$ is
\[
\mathrm{MV}(\itemidx)
= \underbrace{\sensitivity_\creator\, v^*}_{\text{direct creator channel}}
+ \underbrace{\eta_{\mathrm{e}}}_{\text{the view itself}}
+ \underbrace{q'_\itemidx(\budget)\,\Vorg\bigl(\eta_{\mathrm{o}}
    + \sensitivity_\creator v^*\bigr)}_{\text{take-up channel}},
\]
where $\eta_{\mathrm{o}}, \eta_{\mathrm{e}}$ are gross per-view consumption
values of organic and exploration delivery,
$v^* = (\eta_{\mathrm{o}}\bar q\Vorg + \eta_{\mathrm{e}}\bar\budget)/(1-\repro)$
is the total consumption value of one submission, and $\bar q, \bar\budget$
are pool averages.
\end{proposition}
\noindent\emph{Proof sketch.} A marginal view raises total views by
$1 + q'_\itemidx\Vorg$ and induces $\sensitivity_\creator(1 + q'_\itemidx\Vorg)$
submissions, each worth $v^*$, which already sums the descendant chain
(\appref{app:dynamics}).

The take-up term multiplies a small probability change by a large reward. One
extra view barely changes the probability of organic take-up, but conditional
on take-up, the item receives $\Vorg$ additional organic views in expectation.
Two readings of \Cref{prop:gradient} matter later:\par\smallskip

\textbf{(1) Budget scale and ranking.} At the resolution identified here, a
pool-level creator sensitivity re-prices the budget's shadow value but supplies
no new within-pool ranking signal. Creator-term re-ranking requires variation
in $\sensitivity_\creator$ across content items, which none of our designs identifies.
The proposition shows where that heterogeneity would enter. The view-itself term
has no item subscript and drops from the allocator's comparison. The varying
terms are proportional to $q'_\itemidx$ and $\sensitivity_\creator$. Organic
reach $\Vorg$ weights take-up and enters $v^*$, which values induced supply. A
view is worth more near the take-up margin, where $q'_\itemidx$ is
largest, and when conditional organic reach is larger. The latter can
represent demand per unit supply, coverage, diversity, or freshness. When
organic delivery already reaches nearly every item, $q'_\itemidx$ approaches
zero and an extra view buys little beyond itself.

\textbf{(2) Sign.} A randomized contrast measures \emph{net} of the displaced
impression, so a negative measured direct effect is consistent with
$\eta_{\mathrm{e}} > 0$ (\Cref{rem:scope}). Subtracting the same opportunity
cost from every candidate leaves their ordering unchanged, but varying that
cost across candidates can change the ranking.

\subsection{What a Measurement Can See}
\label{sec:measurement}

\Cref{prop:gradient}'s take-up and creator channels are deferred, and this
subsection explains why they are hard to observe. The creator channel arrives a posting cycle late and
compounds over the supply loop. The take-up channel pays off only through
organic distribution that the corpus has yet to deliver. We
analyze the extreme member of the intervention family, switching exploration
off entirely, because it perturbs the corpus stock with the largest possible
asymptotic gap. Any milder test, such as a budget change or parameter tweak,
moves the same stock at the same turnover toward a smaller gap. The switch-off
case is a best-case benchmark for what a fixed horizon can express.
Consumption quantities such as $E_t$ enter below only as observables of the
corpus stock $\corpus_t$, which is not directly observed.

\bpara{Multiplier and horizon.}
For a feedback loop with eigenvalue $\lambda\in(0,1)$, define its equilibrium
multiplier $A(\lambda) \triangleq 1/(1-\lambda)$, the steady-state response to
one unit of input added each period. Its
\emph{relaxation time} is
$\tau(\lambda) \triangleq 1/\ln(1/\lambda)$, so a switch decays as $\lambda^t$.
Both grow without bound as $\lambda$ approaches $1$: stronger feedback also
means slower forgetting. The timing results concern the transient after a
switch, so production need not remain at equilibrium.
The dynamics~\eqref{eq:dynamics} contain two such modes. The supply mode
has eigenvalue $\repro$, the number of posts induced per post, and carries the
comparatively fast posting multiplier. The corpus mode has eigenvalue
$1-\turnover$, its per-period retention, and sets the experiment's clock.
Beyond the immediate slot value, viewer-side effects are corpus-mediated. The
model locates them in time. A short window
reveals only a fraction of a slow corpus effect even without noise.

\begin{theorem}[Horizon blindness]
\label{thm:horizon}
Disable exploration at $t=0$ from equilibrium and write $E^*$ for pre-switch
equilibrium consumption. With consumption affine in the corpus as above, the
gap decomposes as
$E^*-E_t = \Delta_{\mathrm{direct}} + \Delta_{\mathrm{corpus}}\,F(t)$, the direct
term present from the first period. If the corpus retains a fraction
$1-\turnover$ per period and its post-switch inflow stays at or above its
eventual level, then
\[
  F(t) \;\le\; 1-(1-\turnover)^t \;\le\; \turnover t \qquad \text{for every integer } t \ge 1.
\]
\end{theorem}
\noindent\emph{Proof sketch.} Among inflow paths satisfying this condition, an
immediate jump to the eventual post-switch inflow maximizes the corpus gap at every date. Iterating the
stock recursion then gives the normalized gap
$1-(1-\turnover)^t$. Bernoulli's inequality gives the second bound.
The bound is the operational result. At horizon $t$, at most a
$\turnover t$ fraction of the eventual corpus gap has materialized, even
without noise. The linear model
locates the delay. If take-up changes immediately, the first
period expresses a fraction $\turnover$. If corpus inflow changes only after
posting responds, onset is quadratic. When the supply response settles
before the corpus has appreciably turned over, the gap thereafter follows
an approximately linear ramp whose
extrapolated delay lies between zero and one post-switch supply-loop multiplier
$A(\reprooff)=1/(1-\reprooff)$, where $\reprooff$ is the post-switch loop gain.
The delayed endpoint attains that upper limit.
The closed-form path, including mixtures of the two cases, is in
\appref{app:dynamics}. This timing reconciles mixed or negative short
viewer-side readings with a valuable mechanism: the corpus-mediated demand
value arrives on the slow eigenvalue, after the window closes.

Equilibrium gain and relaxation time coincide to within one period. A
coverage target multiplies that time only by its logarithm.

\begin{corollary}[Value and measurability duality]
\label{cor:duality}
Consider a stable nonnegative ecosystem $x_{t+1} = M x_t + u$ and a
feedback loop with real eigenvalue $\lambda \in (0,1)$.
\begin{enumerate}[nosep,leftmargin=1.6em,label=(\roman*)]
  \item the equilibrium multiplier $A(\lambda)$ and relaxation time
  $\tau(\lambda)$ coincide to within one period, so
  $\tau(\lambda) \le A(\lambda) \le \tau(\lambda)+1$. Observing a
  $(1-\varepsilon)$ fraction of a switch requires the horizon
  $t_\varepsilon = \ln(1/\varepsilon)\,\tau(\lambda)$.
  \item consequently an experiment of length $t$ reads the partial sum
  $\sum_{k<t}\lambda^{k} \le t$, whatever the true multiplier
  $A(\lambda)$.
\end{enumerate}
Within the linear model, no parameter regime of such a loop combines large equilibrium value with a short measurement horizon.
\end{corollary}
\noindent Part (ii) is the operational form (full proof in \appref{app:dynamics}).
What a test can read is bounded by its own duration, so a gap settling at
$\Delta_\infty$ reads early as $\Delta_\infty\turnover t$. The multiplier is
achieved \emph{by} the slowness. The corpus stock instantiates the corollary
at eigenvalue $1-\turnover$. Its gain is $1/\turnover$, and its relaxation
time lies within one period of that value. The more evergreen the corpus, the
more stock a unit of inflow buys and the longer any test must run to see a
fixed fraction of it.

Adding users does not change that clock. It only sharpens the reading of
wherever the clock stands. \appref{app:illustrations} works these results
at illustrative magnitudes.

\begin{proposition}[Sample size cannot buy horizon]
\label{prop:detection}
Consider an on/off experiment with $N$ users per arm on a corpus-stock
metric with asymptotic gap $\Delta_\infty$, read at horizon $t$ in the early
regime $\turnover t \ll 1$.
(i) The expressed fraction
$\Delta(t)/\Delta_\infty \approx \turnover t$ is independent of $N$.
Larger samples estimate the same attenuated number more precisely.
(ii) Even mere detection of the attenuated gap is horizon-bound. With
per-user noise scale $\sigma$ and detection level $z$, a snapshot analysis
first detects at
$t_{\det} = \frac{z\sigma\sqrt{2/N}}{\Delta_\infty \turnover}
\propto N^{-1/2}$,
and a cumulative analysis with independent per-period noise at
$t_{\det} = \bigl(\tfrac{2\sqrt{2}\,z\sigma}
{\Delta_\infty \turnover \sqrt{N}}\bigr)^{2/3} \propto N^{-1/3}$.
\end{proposition}
\begin{proof}[Proof sketch]
(i) follows from \Cref{thm:horizon}, whose statement contains no $N$. For (ii), equate the
gap $\Delta_\infty\turnover t$ (snapshot) or its running mean (cumulative,
standard error $\sigma\sqrt{2/(Nt)}$) to $z$ times its standard error and
solve for $t$.
\end{proof}

Detecting a nonzero finite-horizon effect is easier than estimating its
asymptote.

\begin{theorem}[Estimation floor]
\label{thm:curvature}
Let $y_t$ be the observed between-arm contrast in a corpus-stock metric at
horizon $t$. For $t=1,\ldots,T$, suppose
\[
  y_t \;=\; c+\Delta_\infty\bigl(1-e^{-\turnover t}\bigr)+\varepsilon_t,
  \qquad \varepsilon_t \stackrel{\mathrm{iid}}{\sim} N(0,\,s^2),
\]
with level $c$ and rate $\turnover$ unknown and
$s^2 = 2\sigma_{\mathrm{eff}}^2/N$ for $N$ users per arm, where
$\sigma_{\mathrm{eff}}$ is the effective per-user transient-noise scale. Define the
\emph{curvature signal-to-noise ratio}
\[
  \mathcal{C} \;=\; \frac{\Delta_\infty\,\turnover^{2}\sqrt{S_4}}{2s},
  \qquad S_4 = \textstyle\sum_{t \le T} t^{4}.
\]
Whenever $\mathcal{C} \le 1$, every confidence procedure for
$\Delta_\infty$ with uniform coverage $1-\alpha$ over positive amplitudes
and rates returns an infinite upper endpoint with probability at least
$\tfrac12-\alpha$. Raising $N$ moves the crossing $\mathcal{C}=1$ only as
$T \propto N^{-1/5}$, improving to $N^{-1/3}$ once $\turnover$ is estimated
from auxiliary data.
\end{theorem}
\noindent The proof, a two-point total-variation indistinguishability
argument over a matched-product family, appears in \appref{app:floors},
with the panel reduction at \resref{Lemma}{lem:panel} and the formal
statement at \resref{Theorem}{thm:floor}. The construction holds the early slope
$\Delta_\infty\turnover$ fixed while sending $\turnover\to0$ and
$\Delta_\infty\to\infty$. Below the threshold these paths are
statistically indistinguishable. Until the trajectory visibly bends,
monotonicity can still supply an increment lower bound. The slow component
is the corpus stock. The regimes in which exploration is most valuable
are those in which feasible-horizon inference may admit only an increment
lower bound, and \S\ref{sec:probe} uses a horizon-matched endpoint accordingly.

\resref{Figure}{fig:floor-demo} and
\resref{Figure}{fig:supp-two-ends} visualize the estimation floor and delay
family.

\ifextendedbuild
The theorem treats the contrast series as independent draws with a single
noise scale. Panel data earns that form: persistent per-user differences drop
out once the level is profiled, serial correlation is absorbed into
$\sigma_{\mathrm{eff}}$, and only a unit root would change the exponents.

\begin{lemma}[Panel reduction]
\label{lem:panel}
Let per-user outcomes carry persistent user effects of arbitrary variance
$\sigma_\alpha^2$ and stationary transient noise, so that
$\mathrm{Cov}(\eta_s, \eta_t) = \tfrac{2}{N}\bigl[\sigma_\alpha^2 +
\gamma(s-t)\bigr]$. Then:
(i) because $\mathbf{1}$ is an eigenvector of the persistent component,
profiling any nuisance set containing the intercept
$\Delta_{\mathrm{direct}}$ renders every information quantity below
\emph{independent of $\sigma_\alpha^2$}, and persistent effects can only
enlarge the impossibility region of \Cref{thm:floor};
(ii) for AR(1) transient noise with parameter $\rho$ and marginal variance
$\sigma^2$, all statements hold with
$\sigma_{\mathrm{eff}}^2 = \sigma^2(1+\rho)/(1-\rho)$, up to a factor
$1 + O\bigl(\rho/((1-\rho)T)\bigr)$, by the exact tridiagonal form of the
AR(1) inverse;
(iii) stationarity is necessary: under random-walk user drift the $T^5$
information accumulation below degrades to $T^3$, because a unit root
mimics the ramp. Stationarity is diagnosable in logs: the variance of
user-mean-centered outcomes must not grow with the window.
\end{lemma}

\noindent The proof is in \appref{app:floors}. Part (iii) is why
\S\ref{sec:experiments} reports a stationarity check rather than assuming it.
\fi

\bpara{Drift.} A changing environment moves the steady state without changing
timing or identification. \Cref{thm:horizon} uses only the corpus stock's
retention and holds pathwise under time-varying turnover, the gain--relaxation
relation survives because both are monotone in the same $\lambda$, and the
exponents in \Cref{prop:detection} follow from that algebra alone
(\appref{app:dynamics}). Unmodelled drift only widens the region with no
finite upper bound. The multiplier and \Cref{prop:gradient}'s level are
local statements. The probe in \S\ref{sec:probe} imports drift from the
concurrent ablation. Each design contrasts arms at a common date, so
shared changes in viewer taste or the corpus drop out.

\subsection{What Each Design Identifies}
\label{sec:designs}

Each randomization holds a different part of the ecosystem in common between
its arms. What is common cancels.

\begin{proposition}[What each randomization design identifies]
\label{prop:identification}
In the model~\eqref{eq:dynamics}, suppose an experiment holds a fraction $p$
of one side of the market at the perturbed condition (the floored arm).
Treat immediate within-feed substitution as part of the direct channel, and
suppose any remaining change in organic consumption is mediated by the shared
corpus state.
The following statements hold.
\begin{enumerate}[nosep,leftmargin=1.6em,label=(\roman*)]
  \item \textbf{Viewer-side.} Disabling exploration delivery in enrolled
  users' feeds identifies the direct channel exactly at every horizon and
  arm size. Corpus-mediated consumption is common to both arms and cancels.
  \item \textbf{Creator-side.} Holding enrolled creators' content at the
  floor identifies the equilibrium supply response
  $\subs_0\bigl[(1-\repro)^{-1} - (1-\repro_{\mathrm{f}})^{-1}\bigr]$
  \emph{at the perturbed aggregate}, up to a capacity-release term
  $\epsilon(p)$. Here $\repro_{\mathrm{f}}$ is the floored arm's loop gain.
  The term $\epsilon(p)$ is nonnegative because flooring can only raise delivery in
  the non-floored arm relative to full rollout. It scales with the floored
  market share and is absent from the co-diverted design, where an isolated
  submarket exchanges no capacity with the platform. Whatever consumption the
  induced corpus generates is common to both arms and therefore cancels from the contrast.
  \item \textbf{Co-diverted.} Isolating matched creator and viewer
  submarkets retains the corpus-mediated component in the contrast,
  attenuated by the general $F(t)$ of \Cref{thm:horizon}. This differs from
  the one-sided cancellation of (i)--(ii). Flooring the submarket's own creators lowers
  its own supply, so the cell sits toward the theorem's delayed end.
\end{enumerate}
In particular, the corpus channel enters no one-sided contrast at any
horizon or sample size. It enters only under co-diversion and expresses a
fixed fraction at horizons of order $1/\turnover$.
\end{proposition}
\noindent For (i)--(ii), subtracting arm means cancels the common $\corpus_t$.
Under co-diversion each cell evolves its own $\corpus_t$, so subtraction
retains their stock gap. The capacity-release sign in (ii) follows because
flooring enrolled traffic frees nonnegative capacity for the remaining pool.
\appref{app:dynamics} gives the full algebra.

These designs identify population aggregates. Per-item $q'_\itemidx$ and
$\sensitivity_\creator$ remain unidentified. An external turnover estimate improves corpus
estimation from $N^{-1/5}$ to $N^{-1/3}$.

\begin{table}[t]
\centering
\caption{Identification map for the four randomized designs. The final
column gives the sample-size law. For the fast channels it sharpens the
estimate at a fixed horizon. For the corpus rows it sets the required
horizon.}
\label{tab:design-map}
\footnotesize
\setlength{\tabcolsep}{3.5pt}
\begin{tabular}{@{}llll@{}}
\toprule
Target & Design & Required horizon & Rate in $N$ \\
\midrule
Direct effect & Viewer ablation & Immediate & $N^{-1/2}$ \\
Supply response & Creator ablation & Supply loop $\tau(\repro)$ & $N^{-1/2}$ \\
Allocation response & Budget reallocation & Supply loop $\tau(\repro)$ & $N^{-1/2}$ \\
Corpus ramp & Co-diversion & Early ramp & $N^{-1/3}$ \\
Corpus asymptote & Co-diversion & Curvature threshold & $N^{-1/5}$ \\
\bottomrule
\end{tabular}
\end{table}

\section{Testing the Identification Map}
\label{sec:experiments}

The four experiments identify complementary channels. Each tests the channel
and time path assigned to it by
\Cref{tab:design-map}. Two long-running
one-sided ablations identify the direct link and the supply response in sign,
although shared capacity can overstate the latter under full rollout. Two
co-diverted experiments test budget reallocation
(\S\ref{sec:reallocation}) and the delayed corpus response
(\S\ref{sec:probe}), which one-sided designs exclude by construction.
In both ablations, randomization and analysis units coincide, so effects are
differences of window-aggregated means at the randomized-unit level, with
unit-level standard errors and no clustering. Enrollment is fixed at randomization. Where used,
variance reduction regresses outcomes on their pre-experiment
values~\citep{deng2013cuped}. The creator first stage is mechanical. Exploration
views per video rise by the budget-to-floor gap (levels withheld). Arm shares
agree within hundredths of a percentage point, with no sample-ratio mismatch. We
report effect sizes with $95\%$ intervals. Tabulated rows are a subset of each analysis's full outcome
family. The probe reports the endpoints specified in its experiment design.
Each co-diverted experiment realizes one pair of matched submarkets, so its
intervals condition on that pair.

\subsection{Qualitative Predictions}
\label{sec:simulation}

\begin{figure*}[t]
  \centering
  \includegraphics[width=0.87\textwidth]{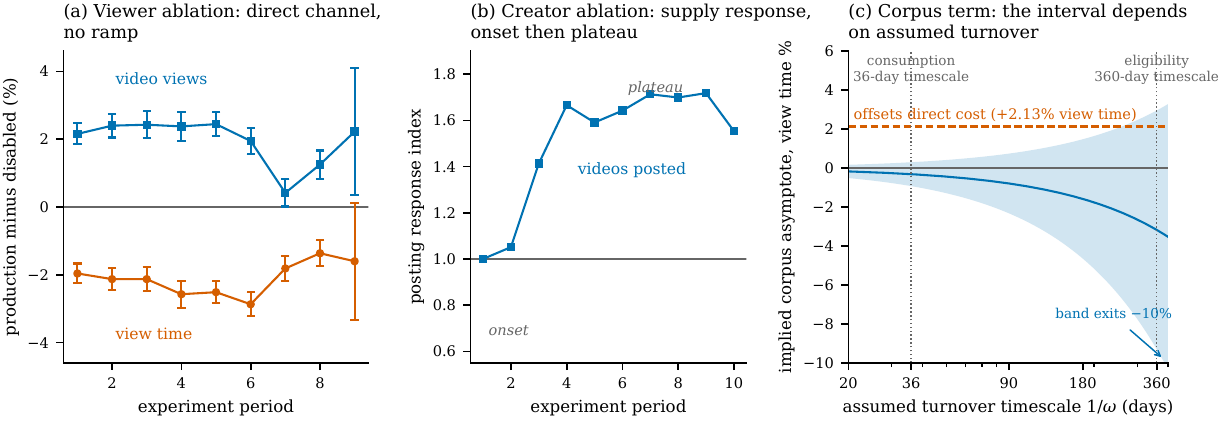}
  \caption{Observed time paths and turnover sensitivity. \textbf{(a)} The year-long viewer contrast
  appears within weeks and stays level until a common operating-point shift,
  matching the direct-step prediction. Intervals are approximate.
  \textbf{(b)} The eight-month posting response, indexed to its first measured
  period, is descriptive and follows onset with a plateau.
  \textbf{(c)} The corpus-asymptote conclusion depends on the assumed turnover. The
  36-day case rules out offsetting either direct view-time cost (the ablation's
  $-2.13\%$, drawn dashed, and the probe's own $-1.02\%$ step), whereas the
  360-day case admits either sign. Panels (a)--(b) use finer analysis periods
  than the \suppname's series. Their shapes agree (\appref{app:timepath}).}
  \label{fig:observed-evidence}
\end{figure*}

The theory implies three signatures independent of effect size. They are a direct step, a
supply response that can build after onset, and a corpus ramp as isolated
stocks diverge. Cumulative and non-overlapping period estimates reveal whether
a gap ramps or stays level. One-sided ablations reach the first two responses.
Only co-diversion can express the corpus ramp. \Cref{fig:observed-evidence}
shows the two observed time paths and the turnover sensitivity below.

\subsection{Viewer Ablation of the Direct Channel}
\label{sec:viewer-ablation}

The viewer ablation ran for over a year, enrolling eight percent of users
randomized between production exploration and full disablement.
Creators are common to both arms, so each arm draws from the same corpus
supplied by the surrounding creator population. The contrast isolates the
\emph{direct channel}, which is the effect of exploration delivery in enrolled
viewers' feeds after within-feed substitution. It excludes corpus effects. Its
length separates persistent response from launch novelty and spans a full
seasonal cycle.

\begin{table}[t]
\centering
\caption{Randomized effects at the tested operating points. Relative arm
differences have 95\% CIs. Positive values favor production in Panels A--B
and reallocation in Panel C. The panels isolate a mixed direct trade-off, a
signed supply response, and a budget-matched participation gain.}
\label{tab:ablations}
\normalsize
\renewcommand{\arraystretch}{0.90}
\setlength{\tabcolsep}{4pt}
\begin{tabular*}{0.96\columnwidth}{@{\extracolsep{\fill}}lrr@{}}
\toprule
Metric & Rel.\ change (\%) & 95\% CI \\
\midrule
\multicolumn{3}{@{}l}{\textbf{Panel A: Viewer ablation} (\S\ref{sec:viewer-ablation})} \\
Video views      & $+1.74$ & $[+1.57, +1.90]$ \\
View time        & $-2.13$ & $[-2.28, -1.98]$ \\
Views under $1.5\,s$ & $+7.49$ & $[+7.28, +7.70]$ \\
Video favorites  & $-2.32$ & $[-2.72, -1.92]$ \\
\midrule
\multicolumn{3}{@{}l}{\textbf{Panel B: Creator ablation} (\S\ref{sec:creator-ablation})} \\
Videos posted per creator & $+8.55$ & $[+7.14, +9.96]$ \\
Creators posting at least once & $+7.10$ & $[+6.60, +7.59]$ \\
\midrule
\multicolumn{3}{@{}l}{\textbf{Panel C: Budget-matched reallocation} (\S\ref{sec:reallocation})} \\
Creators posting at least once & $+0.77$ & $[+0.35, +1.19]$ \\
Videos posted per creator & $+0.92$ & $[+0.05, +1.79]$ \\
View time        & $+0.01$ & $[-0.14, +0.16]$ \\
Video views      & $-0.06$ & $[-0.21, +0.09]$ \\
Video favorites  & $+0.15$ & $[-0.44, +0.74]$ \\
\bottomrule
\end{tabular*}
\end{table}

Table~\ref{tab:ablations} (Panel A) reports a mixed direct effect. Video views
rise 1.74\% while view time falls 2.13\%. Views under $1.5\,s$ rise $7.5\%$,
shifting the view mix toward brief views. Favorites move with view time
($-2.3\%$). Exploration increases viewing starts while reducing viewing duration at
this operating point. This is the direct-feed component of the allocation problem.
\Cref{prop:gradient} also requires the creator response.

The cumulative and non-overlapping period series in \appref{app:timepath}
confirm the reading. The gap is close to full size within weeks. Thereafter
the period contrast stays level across most of the window and then attenuates,
tracking a decline in exploration's share of the feed. Cumulative video views
move from $+2.02\%$ early in the window to $+2.32\%$ at the peak and
$+1.74\%$ at the end. Immediacy without a sustained ramp is the direct
channel's signature under the reading rule of \S\ref{sec:simulation}. The
shared corpus contributes zero to this contrast.

The concurrent creator ablation randomizes a different population, yet its
consumption changes sign at the same calendar boundary ($+1.11\%$ on video
views just before it), as shown in \appref{app:timepath}. Because the experiments began
months apart, the same calendar-date break indicates a common operating-point
shift. Since both designs exclude the corpus term
(\Cref{prop:identification}(i)--(ii)), this concurrent movement also supplies the
drift estimate imported by \S\ref{sec:probe}.

\subsection{Creator Ablation of Content Supply}
\label{sec:creator-ablation}

The creator ablation ran for eight months, enrolling eight percent of the
platform's creators, with about four percent in each arm.
This share also sizes the interference term, with $p/(1-p) \approx 0.04$.
Creators in the \emph{floored} arm receive a minimal exploration budget per
video, and their videos remain eligible for organic retrieval. Creators in the
\emph{production} arm receive production exploration budgets (absolute levels withheld for
confidentiality). This two-dose comparison
falls short of a strict on/off but measures the supply response directly.
Only the budget above the floor differs between arms.

Across eight months, exploration raises videos posted per creator by 8.55\%
(Table~\ref{tab:ablations}, Panel B). It also raises the number of creators
posting at least once by 7.10\%. The two outcomes identify aggregate and
extensive-margin supply responses.
Unlike randomized traffic grants lasting a few weeks, this contrast compares
the standing production mechanism with a minimal floor over eight months.

The effect on videos posted per creator is $+5.9\%\pm3.4$ ($95\%$) in the first
measured period, below the $+8.55\%$ whole-window average. Together with the
later plateau in \Cref{fig:observed-evidence}b, this pattern is consistent with
an onset period. The increase in creators posting at least once locates part of
the response on the extensive margin.
Because the two creator arms compete in a shared marketplace,
\Cref{prop:identification}(ii) implies capacity-release inflation at order
$p/(1-p)$, so the contrast upper-bounds the full-rollout effect at this operating point
(\appref{app:dynamics}). The co-diverted probe later provides a descriptive
comparison without that shared-capacity term. Recovering the guarantee's total value would
require a zero-exploration creator arm.

\subsection{Budget-Matched Reallocation of Exposure}
\label{sec:reallocation}

The ablations identify the effects of exploration presence. \Cref{prop:gradient}
also characterizes fixed-budget allocation. A two-week co-diverted experiment holds nominal budget fixed while
spreading exposure across more videos in matched cells. Its design fixes the
arm contrast, horizon, and endpoints. By concavity, the
marginal posting response is larger at lower exposure. Promotion thresholds implement the
shift. A penultimate-day spend diagnostic puts realized per-creator budgets
within about $3\%$ across arms, with the reallocated arm slightly lower, so
any residual dose imbalance biases its posting response downward.
The experiment identifies the randomized reallocation contrast and leaves
pure breadth elasticity open. The horizon
targets fast supply. Creators posting at least once increase by $0.77\%$, and
videos posted per creator increase by $0.92\%$. Every viewer-side interval includes zero
(Table~\ref{tab:ablations}, Panel C). This is the predicted signature of higher
creator participation without a detectable short-run viewer cost
(\appref{app:matched-budget}).

\subsection{Co-Diverted Probe of the Corpus Response}
\label{sec:probe}

The probe co-diverts viewers and creators into isolated production and floored
submarkets~\citep{masoero2021multiple,brennan2023symbiosis}. Consumption uses
viewer-level means. Posting uses creator-level means. With one partition per
arm, intervals condition on these submarkets. Matched isolation cancels the
common consumption cost while letting cells approach different steady
states~\citep{shen2026isolation}.

We test video views and view time for a step followed by a delayed ramp. Video
favorites per delivered view provide a quality endpoint sensitive to composition.
The experiment runs for three weeks from switch-on. The corpus estimator omits
the first seven days because videos already in their exploration windows are only
partially exposed to the intervention.

The model predicts an immediate step plus a delayed corpus ramp at rate
$\turnover$, while the concurrent one-sided ablation stays flat
(\Cref{thm:horizon}). Because early data identify only the slope
$\Delta_{\mathrm{corpus}}\,\allowbreak\turnover$, we report asymptote intervals
conditional on turnover and a shape-robust increment.
The 36-day timescale is the measured kernel's single-rate equivalent. The 360-day
timescale is the eligibility horizon. They are fixed-turnover sensitivity cases.
Kernel calibration and estimator details are in \appref{app:estimator}.
Platform policy caps matched-submarket isolation at three weeks because its
ecosystem cost grows with duration~\citep{shen2026isolation}. At this horizon,
\Cref{thm:curvature} predicts an identified direct step but a potentially
unsigned corpus asymptote (\resref{Figure}{fig:floor-demo}).

\begin{table}[t]
\centering
\caption{Co-diverted three-week readout at the calibrated estimation floor.
Entries are relative changes with 95\% CIs. Panel B gives intervals for the
set-identified corpus asymptote under two turnover timescales. Panel C gives feasible-horizon
gross flow; its implied whole-cell contribution is a derived range. Posting is quoted as estimate $\pm$ margin. Panels C--D retain
the inferential qualifications stated in text.}
\label{tab:probe}
\small
\renewcommand{\arraystretch}{0.90}
\setlength{\tabcolsep}{3pt}
\begin{tabular*}{0.96\columnwidth}{@{\extracolsep{\fill}}lr@{}}
\toprule
Quantity & Estimate \\
\midrule
\multicolumn{2}{@{}l}{\textbf{Panel A: Direct step}} \\
Video views & $+0.43\ [+0.30, +0.57]$ \\
View time & $-1.02\ [-1.14, -0.89]$ \\
Video favorites / delivered view & $-0.87\ [-1.21, -0.53]$ \\
\midrule
\multicolumn{2}{@{}l}{\textbf{Panel B: Corpus asymptote by turnover timescale}} \\
Video views, 36-day timescale & $[-0.28, +0.70]$ \\
View time, 36-day timescale & $[-0.93, +0.30]$ \\
Video views, 360-day timescale & $[-2.77, +7.04]$ \\
View time, 360-day timescale & $[-9.30, +2.95]$ \\
\midrule
\multicolumn{2}{@{}l}{\textbf{Panel C: Feasible-horizon corpus flow}} \\
\shortstack[l]{Organic view time in cohort slice\\Implied whole-cell contribution} &
  \shortstack[r]{$+37.9\ [+37.1, +38.7]$\\$+0.1$--$+0.2$ pp} \\
\midrule
\multicolumn{2}{@{}l}{\textbf{Panel D: Videos posted per creator at a matched 21-day horizon}} \\
In-cell design & $+0.63 \pm 0.44$ \\
One-sided design & $+2.79 \pm 0.89$ \\
\bottomrule
\end{tabular*}
\end{table}

\Cref{tab:probe} collects the readout. The direct step repeats the viewer
ablation's signs and is smaller, consistent with the probe's later operating
point. Video favorites per delivered view confound reception with composition.
The estimates use an equal-length pre-period difference because video views and
favorites were imbalanced before the switch.

\begingroup
\setlength{\emergencystretch}{2em}
The corpus readout lands at the predicted estimation floor. Without fixing a
turnover timescale, neither the ramp fit nor the shape-robust increment yields
a finite two-sided upper bound on the corpus asymptote
(\appref{app:estimator}). Under the 36-day turnover timescale,
the auxiliary analysis rules out a corpus benefit large enough to offset the
direct view-time cost. Under the 360-day timescale, either sign remains possible
(\Cref{tab:probe}, Panel B).
The aggregate corpus asymptote remains unsigned at three weeks.
\par
\endgroup

\bpara{Feasible-horizon corpus flow.}
Before the aggregate trajectory bends, the theory targets the effect expressed
at horizon $t$. An extrapolated asymptote is premature. A cohort decomposition
measures feasible-horizon gross flow among videos posted during the experiment
and later served organically after their exploration windows. By construction,
exploration can no longer deliver these videos (\appref{app:cohortledger}). In matched
cells, organic view time is $37.9\%$ higher in a slice representing
$0.3$--$0.5\%$ of cell consumption, equivalent to $0.1$--$0.2$ percentage
points of the whole-cell contrast (\Cref{tab:probe}). Videos posted before the
experiment show no detectable difference after the first seven days. The pattern first appeared in a $1\%$
discovery sample. Because the full-population calculation reuses those users,
we retain its exploratory status. It measures gross corpus flow by the
feasible horizon. The net asymptote after displacement remains unresolved
(\appref{app:cohortledger}).

\bpara{In-cell supply.}
With a comparable exploration first stage and the same horizon and
instrument, in-cell posting is roughly a quarter of the one-sided response
(\Cref{tab:probe}, Panel D). The ratio is descriptive because the
populations, seasons, reach, and exposure composition differ. It leaves
$\epsilon(p)$ and the rollout asymptote unidentified.

\subsection{What the Four Experiments Establish}
\label{sec:virtuous-cycle}

The experiments form an identification sequence across distinct estimands.
The measured timescales match the model's assignment of roles: the posting
response plateaus within a few measured periods
(\Cref{fig:observed-evidence}b), while the two candidate corpus timescales range
from comparable to roughly an order of magnitude longer than the posting
response (\S\ref{sec:probe}).
\Cref{tab:claims} records the findings and nearest limits.

\begin{table}[b]
\centering
\caption{What the four experiments establish. Signs apply at the realized
horizon. Co-diverted rows condition on the realized submarkets.}
\label{tab:claims}
\small
\renewcommand{\arraystretch}{1.08}
\setlength{\tabcolsep}{3pt}
\begin{tabular}{@{}>{\raggedright\arraybackslash}p{0.18\columnwidth}
  >{\raggedright\arraybackslash}p{0.39\columnwidth}
  >{\raggedright\arraybackslash}p{0.35\columnwidth}@{}}
\toprule
Experiment & Identified finding & Leaves open \\
\midrule
Viewer ablation & Direct feed trade-off: video views $+$, time $-$ & Shared-corpus effect \\
\addlinespace[2.5pt]
Creator ablation & Supply response: posting $+$ & Full-rollout magnitude and corpus value \\
\addlinespace[2.5pt]
Budget reallocation & Reallocation contrast: participation $+$; viewer CIs span 0 & Breadth elasticity and long-run corpus value \\
\addlinespace[2.5pt]
Co-diverted & Direct step mixed; exploratory gross cohort flow $+$ & Net corpus asymptote at three weeks \\
\bottomrule
\end{tabular}
\end{table}

\section{Discussion}
\label{sec:discussion}

The creator-side response is consistent with creators on the margin of
continued participation, where modest distribution separates occasional
from regular posting~\citep{bhargava2022creator,qian2024digital}. Organic
delivery accrues mostly to videos that have demonstrated strong early performance, which is
why exploration is the primary channel for most new videos
(\S\ref{sec:intro}) and the platform's direct lever over the long tail. \Cref{prop:gradient} shows where the creator
response enters value accounting. Our designs identify only its gross
aggregate. Per-item sensitivity and the common net-value scale required to
prescribe an objective remain unidentified.

\bpara{Implications for Practice.}
Each result yields a design implication.
\begin{enumerate}[nosep,leftmargin=1.6em]
  \item \textbf{A mixed viewer A/B identifies only the direct trade-off.} A viewer-side
  contrast excludes creator response by construction. On the platform we study,
  production exploration raises videos posted per creator by 8.55\% and creators
  posting at least once by 7.10\%.
  \item \textbf{A neutral result leaves shared-corpus value unresolved.}
  The shared-corpus term cancels from a one-sided viewer contrast. A short
  co-diverted contrast identifies only the finite-horizon increment
  (\Cref{thm:horizon}). The asymptote remains unresolved. Neither result alone
  identifies total value.
  \item \textbf{Estimate the slow factors offline and import them.} Traffic
  cannot rescue the gate's short window (\Cref{prop:detection}). Budget
  policies follow a planning cadence that can incorporate offline estimates
  from past randomized cohorts or quasi-experimental variation. Historical
  data on yield by video age can estimate corpus turnover.
  \item \textbf{Match the design to the channel.} Viewer-side randomization
  measures the direct channel, and creator-side randomization measures the
  aggregate supply response at the tested operating point. Only co-diversion
  retains the corpus channel required to estimate its asymptote
  (\Cref{tab:design-map}). The exploratory
  cohort analysis measures gross corpus flow at a feasible horizon using videos
  posted during the experiment and served organically after their exploration windows
  (\Cref{tab:probe}; \S\ref{sec:probe}).
  \item \textbf{Plan in calendar time.} A tenfold increase in users reduces the required
  horizon by at most a factor of $3.2$. Below the curvature threshold (\Cref{thm:curvature}),
  use the finite-horizon increment and any model-restricted lower bound as the
  reported endpoints.
  Plan horizons from turnover, effect-size, and noise inputs.
  \item \textbf{Treat allocation as a separate experimental margin.} At a
  fixed nominal budget, reallocation raises creator participation with no
  detectable viewer-side change (\S\ref{sec:reallocation}).
\end{enumerate}

\bpara{Limitations.}
The ablations estimate the budget's effects at the tested operating point.
The reallocation experiment tests one direction of allocation there, leaving
the allocation rule unidentified (\S\ref{sec:reallocation}).
Open questions include posting-response persistence, generalizability
beyond one platform, and whether a learned allocation rule improves on the
production rule~\citep{swaminathan2015self,jeon2025epinet}. Nothing
measured here depends on that rule (\S\ref{sec:budget-problem}).

\section{Conclusion}
\label{sec:conclusion}

Content exploration reaches beyond the feed by changing creator participation
and replenishing the shared corpus. Standard one-sided viewer tests identify
only the direct feed channel. Our decomposition separates a delivered view into immediate value,
organic take-up, and induced creator supply. The latter two pay off through the
shared corpus. Horizon blindness bounds what any experiment or the mechanism's
own gate can observe of that deferred value.
Relative to a minimal floor, production exploration raises videos posted per
creator by 8.55\% and creators posting at least once by 7.10\%. At a matched
nominal budget, reallocation raises creator participation with no detectable
short-run viewer-side change. The viewer ablation raises video views while
lowering view time, a mixed direct trade-off. These outcomes use different
units, and total viewer value remains unidentified because one-sided designs
cancel the corpus term. The co-diverted probe reaches that term, but its sign remains
unresolved under both reported turnover scenarios. Evaluating exploration as an
ecosystem intervention requires matching the randomization to the target channel
and planning the horizon from corpus turnover.

\bibliographystyle{ACM-Reference-Format}
\bibliography{references}

\appendix

\noindent\textbf{Appendix roadmap.} The appendices carry the material the
conference submission holds as separate supplementary matter, bound in here
so that this version is self-contained. They follow the order of the sections
they support. \Cref{app:props} derives the ecosystem dynamics of
\S\ref{sec:problem} and proves the information limits stated in
\S\ref{sec:theory}.
\Cref{app:extended} states the results omitted from the main text for length
and specifies the estimator used in the co-diverted analysis.
\Cref{app:timepath} reports the additional empirical analyses behind
\S\ref{sec:experiments}, together with their inferential status.
\Cref{app:illustrations} gives worked numerical illustrations of the theory.

\section{Supporting Propositions and Proofs}
\label{app:props}

This appendix collects the formal statements summarized in
\Cref{sec:dynamics} and \Cref{sec:measurement}. Notation is as in the main
text.

\subsection{Ecosystem dynamics}
\label{app:dynamics}

\ifextendedbuild
\noindent The two results of \S\ref{sec:problem}--\S\ref{sec:theory} that
describe where the ecosystem settles, \Cref{prop:multiplier} and
\Cref{prop:gradient}, are proved here in the order they are stated.
\else
\begin{proposition}[Exploration multiplier]
\label{prop:multiplier}
For $\repro < 1$, the dynamics~\eqref{eq:dynamics} converge to
\[
\subs^* = \frac{\subs_0}{1-\repro},
\qquad
\corpus^* = \frac{q(\budget)}{\turnover}\,\subs^*,
\]
and one additional delivered view yields $\sensitivity/(1-\repro)$
cumulative additional submissions over the ecosystem's response. Every induced
submission attracts views that induce further submissions, producing the multiplier.
\end{proposition}
\begin{proof}
The affine recursion for $\subs_t$ contracts to its unique fixed point
$\subs_0/(1-\repro)$. Substituting into the $\corpus$ recursion gives
$\corpus^*$. A marginal view induces $\sensitivity$ submissions, which
induce $\sensitivity\repro$, then $\sensitivity\repro^2$, summing to
$\sensitivity/(1-\repro)$.
\end{proof}

\fi

\begin{proof}[Proof of \Cref{prop:gradient}]
A marginal view raises the content item's total views by
$1 + q'_\itemidx(\budget)\Vorg$. It does so once directly and again through the
take-up probability, since a taken-up
item receives a further $\Vorg$ organic views. The marginal exploration view
is itself consumed and is worth $\eta_{\mathrm{e}}$. The induced organic
views are worth $\eta_{\mathrm{o}}$ each. Because the model credits all of an
item's views with inducing posting, the same increment induces
$\sensitivity_\creator\bigl(1 + q'_\itemidx(\budget)\Vorg\bigr)$ submissions
from that creator. Each submission is worth $v^*$, which satisfies
$v^* = \eta_{\mathrm{e}}\bar\budget + \eta_{\mathrm{o}}\bar q \Vorg
+ \repro v^*$, so $v^*$ already contains the whole descendant chain and no
further multiplier is applied. Collecting the posting term
$\sensitivity_\creator v^*(1 + q'_\itemidx\Vorg)$ with the consumption term
$\eta_{\mathrm{e}} + q'_\itemidx\Vorg\eta_{\mathrm{o}}$ gives the statement.
\end{proof}

\noindent \Cref{thm:horizon} on horizon blindness and \Cref{cor:duality} on
the value and measurability duality are stated in \Cref{sec:measurement}.
Their proofs are below. The detection-scaling
result (\Cref{prop:detection}) is stated with proof there too.

\paragraph{Scope of the linear model: the discussion behind \Cref{rem:scope}.}
Three effects are deliberately outside the recursions~\eqref{eq:dynamics}.

First, $\Vorg$ is held at its steady-state level, in the budget and in the
corpus, and the two dependences thereby suppressed are both signed. If extra
exploration also improves how a content item performs organically \emph{conditional}
on take-up, the marginal view earns a further $q(\budget)\Vorg'(\budget) \ge 0$,
so \Cref{prop:gradient} understates it and the take-up channel here is the
extensive margin only. If attention is finite, $\Vorg$ instead falls as the
corpus grows, which lowers $\bar V(\budget)$ and hence $\repro$. That is the
negative feedback capacity rationing already produces (\Cref{rem:capacity}),
with the same consequence, an overstated multiplier and untouched timing
statements. The stock-timing bound holds structurally. Since $\Vorg$ enters nowhere in
the corpus recursion, whose inflow $q(\budget)\subs_t$ is counted in content items,
\Cref{thm:horizon} uses only the stock's retention and inflow staying at or
above its eventual level, and is unaffected by any time path of $\Vorg$. Translating that
stock bound to consumption retains the main text's local affine observation
model. Second, the posting response $\creatorresp$ is concave and saturating, a
modeling assumption consistent with the field literature. We linearize this
response only at its current steady state. Together with a positive intercept
and the assumed subcritical slope, concavity gives a unique stable equilibrium.
Its local dynamics are governed by the loop gain
$\repro = \creatorresp(\bar V(\budget))$, and concavity signs
the linearization error. Comparative statics in $\budget$ overstate the
response to budget increases and understate it to decreases, while the
recursion in $\subs$ itself is exact at fixed $\bar V$. Third, attention is
finite. Exploration views displace organic impressions, so a global budget
change induces crowd-out that a linear model ignores. An on/off contrast
between randomized arms measures the equilibrium gap net of this
reallocation. The model also omits the viewer-side loop, in which a richer
corpus draws more viewing and hence more deliverable capacity (the
corpus-to-viewing link, distinct from the exploration-to-viewing link
measured in \Cref{sec:viewer-ablation}). Including it adds a nonnegative
coupling to $M$ and can only raise its spectral radius, hence the multiplier,
so its omission is conservative in the direction of the paper's claims. The same monotonicity covers network effects generally:
further nonnegative feedbacks (views accumulating followers, thicker
corpora matching viewers better) add stocks or couplings to the
nonnegative $M$ and can only raise its spectral radius, so each slows the
slowest component and enlarges the multiplier. The model's role is to explain the
timing and magnitude structure of an on/off contrast. Forecasting levels
from micro-parameters alone is outside its scope.

\paragraph{Fixed capacity is self-damping: the argument behind
\Cref{rem:capacity}.}
Under fixed aggregate capacity $\totalviews$ the pool of $\subs$ content items
shares it, so $\bar V = \totalviews/\subs$ and the induced-submission map is
$\varphi(\subs) = \subs\,\creatorresp(\totalviews/\subs)$, the
\emph{perspective transform} of $\creatorresp$. Write $u = \totalviews/\subs$
for per-item views. Differentiating,
\[
  \varphi'(\subs) \;=\; \creatorresp(u) - u\,\creatorresp'(u).
\]
Three consequences. \emph{(i) Monotone and concave.} Concavity of
$\creatorresp$ with $\creatorresp(0) \ge 0$ gives
$\creatorresp(0) \le \creatorresp(u) - u\,\creatorresp'(u)$, so
$\varphi' \ge \creatorresp(0) \ge 0$; and $\varphi$ inherits concavity from
$\creatorresp$ under the perspective transform. \emph{(ii) Unique globally
stable equilibrium.} $\subs \mapsto \subs_0 + \varphi(\subs)$ is concave and
increasing with $\varphi(0) = 0$, so $\varphi(\subs)/\subs$ is decreasing and
the fixed point is unique; it is stable whenever
$\varphi'(\subs^*) < 1$, which subcriticality already gives since
$\varphi'(\subs^*) < \repro < 1$ by (iii). \emph{(iii) Strictly smaller
gain.} At the same per-item view level $\repro = \creatorresp(u)$, so
\[
  \varphi'(\subs) \;=\; \creatorresp(u) - u\,\creatorresp'(u)
  \;=\; \repro\Bigl(1 - \tfrac{u\,\creatorresp'(u)}{\creatorresp(u)}\Bigr),
\]
whose bracketed subtrahend is the \emph{local elasticity} of the response at
$u$; concavity with $\creatorresp(0)\ge 0$ puts it in $(0,1)$, so
$\varphi'(\subs) < \repro$ strictly. The elasticity is constant only for a
power law: for a saturating response such as the logarithmic one it falls as
the view level rises, so the damping is strongest where per-item views are
thinnest.
Fixed capacity therefore lowers the loop gain and the multiplier
$1/(1-\repro)$. It leaves $\turnover$ unchanged, so \Cref{thm:horizon} and
\Cref{cor:duality} apply unchanged to the corpus stock, and the timing
conclusions hold a fortiori. This is the one negative feedback the paper
accounts for explicitly (\Cref{rem:scope}).

\begin{proof}[Proof of \Cref{cor:duality}]
The geometric sum gives
$\sum_{k\ge0}\lambda^{k} = 1/(1-\lambda) = A(\lambda)$, while setting
$\lambda^{t} = \varepsilon$ gives the expression horizon
$t = \ln(1/\varepsilon)/\ln(1/\lambda) = \ln(1/\varepsilon)\,\tau(\lambda)$.
On $(0,1)$, $1-\lambda \le \ln(1/\lambda) \le (1-\lambda)/\lambda$;
taking reciprocals, the left inequality gives $\tau(\lambda) \le A(\lambda)$
and the right gives
$A(\lambda) - 1 = \lambda/(1-\lambda) \le \tau(\lambda)$.
For part (ii), the partial sum is $(1-\lambda^{t})/(1-\lambda)$, which is bounded by
$t$ for every $\lambda \in (0,1)$ and tends to $t$ as $\lambda \to 1$.
\end{proof}

\paragraph{Full solution behind \Cref{thm:horizon}.}
Write $\subsoff$ and $\corpusoff$ for the post-switch supply and corpus
equilibria, $\Delta s = \subs^* - \subsoff$ for the supply gap, and
$\qoff = q(0)$ for post-switch take-up. The submission recursion solves to
$\subs_t = \subsoff + \Delta s\,(\reprooff)^{t}$. Substituting into the
corpus recursion and solving the driven linear recursion gives, for
$\reprooff \neq 1-\turnover$ (the coincident case is the limit of what
follows, in which the $(1-\turnover)^{t}$ term acquires a factor $t$),
\[
\corpus_t
= \corpusoff
+ \bigl(\corpus^{*} - \corpusoff + \kappa\bigr)(1-\turnover)^{t}
- \kappa\,(\reprooff)^{t},
\qquad
\kappa \triangleq \frac{\qoff\,\Delta s}{1-\turnover-\reprooff},
\]
which satisfies $\corpus_0 = \corpus^{*}$. The expressed fraction of the
corpus gap is therefore exactly
\[
\frac{\corpus^{*}-\corpus_t}{\corpus^{*}-\corpusoff}
= 1-(1-\turnover)^{t}
- \frac{\kappa}{\corpus^{*}-\corpusoff}
  \Bigl[(1-\turnover)^{t}-(\reprooff)^{t}\Bigr].
\]
The corpus gap satisfies $\corpus^{*}-\corpusoff
= \bigl(q(\budget)\,\subs^{*} - \qoff\subsoff\bigr)/\turnover
\ge \qoff\,\Delta s/\turnover$ because $q(\budget) \ge q(0)$. When
$\reprooff < 1-\turnover$, the correction coefficient
$c \triangleq \kappa/(\corpus^{*}-\corpusoff)$ obeys
$0 \le c \le \turnover/(1-\turnover-\reprooff)$,
which is $O(\turnover/(1-\reprooff))$ in the regime
$1-\reprooff \gg \turnover$. That regime is where the closed form is most
readable. At this platform's measured parameters, it is comfortable under the
360-day corpus turnover timescale and
marginal under the 36-day timescale, where $1-\turnover$ approaches $\reprooff$ and $\kappa$ and
the bound on $c$ both grow. The theorem itself is unaffected, since the
paragraph below derives it without \eqref{eq:Fexact}. The corpus-mediated
fraction is, exactly and in discrete time,
\begin{equation}
F(t) \;=\; \bigl[1-(1-\turnover)^{t}\bigr]
\;-\; c\,\bigl[(1-\turnover)^{t}-(\reprooff)^{t}\bigr] .
\label{eq:Fexact}
\end{equation}
All three statements of \Cref{thm:horizon} follow from \eqref{eq:Fexact}
directly, with no continuous-time step. \emph{Bound:} for
$\reprooff \le 1-\turnover$ both $c$ and the second bracket are
nonnegative; in the opposite ordering both change sign together, so the
subtracted product is nonnegative either way and
$F(t) \le 1-(1-\turnover)^{t} \le \turnover t$ for every $t$ and every
admissible $c$.

\paragraph{The bound without the linear model.}
That derivation runs through \eqref{eq:Fexact} and so through the linearized
supply loop, but the bound needs neither. Let the corpus be any stock with
per-period retention $1-\turnover$ and inflow $I_t$, so
$\corpus_{t+1} = (1-\turnover)\corpus_t + I_t$, and let $I_{\infty}$ be the
post-switch inflow with $\corpus_{\infty} = I_{\infty}/\turnover$. Writing
$D_t = \corpus_t - \corpus_{\infty}$ and subtracting the fixed-point identity
from the recursion,
\[
  D_{t+1} \;=\; (1-\turnover)D_t + (I_t - I_{\infty}).
\]
If the inflow stays at or above its eventual level, $I_t \ge I_{\infty}$, then
$D_{t+1} \ge (1-\turnover)D_t$, so $D_t \ge (1-\turnover)^{t}D_0$, and for a
switch-off ($D_0 > 0$) the expressed fraction obeys
\[
  F(t) \;=\; 1 - D_t/D_0 \;\le\; 1-(1-\turnover)^{t} \;\le\; \turnover t .
\]
Conditional on $I_t \ge I_{\infty}$, no property of $\creatorresp$ is used.
The bound therefore covers nonlinear creator responses and supply loops. Only
the corpus's constant fractional turnover and the stated inflow condition enter.
Both are mild here. An increasing concave self-map with positive intercept and
slope below one at its fixed point has a unique fixed point, and iterates
converge to it monotonically (\Cref{rem:scope}), which gives
$I_t \ge I_{\infty}$; and the
bound is tight, with equality exactly when the inflow steps to $I_{\infty}$ at
once. Turnover may also vary: with rates $\turnover_s$ the same argument gives
$F(t) \le 1 - \prod_{s<t}(1-\turnover_s) \le \sum_{s<t}\turnover_s$, which is
what a drifting platform satisfies. The argument also allows stochastic inflow
and turnover. It is an inequality on $D_t$ realization by realization, so the
bound holds \emph{pathwise} and the realized gap never exceeds it. A finite-sample
\emph{estimate} of the gap can,
by sampling error alone, which is why detection is treated separately in
\Cref{sec:measurement}. With constant turnover, exceeding the bound requires
an inflow path that falls below its eventual level. The monotone concave loop
of \S\ref{sec:dynamics} cannot produce that path. The shape does require constant
coefficients. The two-term $F(t)$ of \eqref{eq:Fexact}, the delay range, and
\Cref{cor:duality} all need the constant coefficients, and
\Cref{prop:gradient} is a gradient and so is local by construction. \emph{The two ends:} at $t=1$, \eqref{eq:Fexact} gives
exactly $F(1) = \turnover - c\,(1-\turnover-\reprooff)$, so $c$ is precisely
what is subtracted from the corpus's own first-period response. At $c=0$
(take-up falls independently of any change in submissions, e.g.\ the
exploration-window signal itself stops sustaining organic take-up at the
\emph{same} supply level) $F(1)=\turnover$: the pure corpus response,
undelayed. At the upper bound $c=\turnover/(1-\turnover-\reprooff)$, attained
when the switch moves the supply loop alone ($q(\budget)=q(0)$, take-up rate
unchanged, only submission volume falls) $F(1)=0$ exactly: this is the case
of maximal delay, and the one the theorem's sharper reading names.
\emph{Onset at that upper bound:} the first period expresses nothing at
all, which is the mechanical reading of the delay: the corpus cannot
begin to thin until the submissions feeding it have themselves fallen,
and the onset is therefore \emph{quadratic}. Expanding \eqref{eq:Fexact} to
first order in $\turnover$ at $c = c_{\max}$ gives
\[
  F(t) \;=\; \turnover\Bigl[t - \frac{1-(\reprooff)^{t}}{1-\reprooff}\Bigr]
  \;+\; O\bigl((\turnover t)^{2}\bigr),
\]
which vanishes at $t=1$, as $F(1)=0$ requires, and grows as
$\turnover(1-\reprooff)\,t(t-1)/2$ while $(1-\reprooff)t \ll 1$. We state the
discrete form. The continuous $\turnover t^2/(2\tau(\reprooff))$ requires both
$t \gg 1$ and $\reprooff \to 1$. At $\turnover=0.005$,
$\reprooff=0.9$ it overstates $F(2)$ by a factor of two and $F(5)$ by half,
even inside the regime $t \ll \tau(\reprooff)$ where the continuous form is
often invoked.
\emph{Ramp:} once $(\reprooff)^{t}$ is negligible and $\turnover t \ll 1$,
substituting $(1-\turnover)^{t} = 1-\turnover t + O((\turnover t)^2)$ gives
$F(t) = (1+c)\turnover t - c + O((\turnover t)^2)$, a line crossing zero at
$t_0 = c/[(1+c)\turnover]$
exactly, for every admissible $c$. The common $c/\turnover$ expression omits the
$(1+c)$ factor in the slope. At the upper bound
$c=\turnover/(1-\turnover-\reprooff)$, $1+c$ simplifies to
$(1-\reprooff)/(1-\turnover-\reprooff)$ and $t_0$ to $1/(1-\reprooff) = A(\reprooff)$
\emph{exactly}, with no approximation beyond the linearization already made.
The inferred delay is one supply-loop multiplier whatever $\turnover$ is
relative to $1-\reprooff$. The intercept is an extrapolation. At
$\turnover=0.005$, $\reprooff=0.9$ the exact $F(A(\reprooff)) = 0.017$,
almost exactly the fast term $c\,(\reprooff)^{A(\reprooff)} = 0.018$ that the
ramp discards. At $c=c_{\max}$ the only zeros of $F$ are $t=0$ and $t=1$. Only the
ramp's \emph{slope} away from that crossing,
$\turnover(1-\reprooff)/(1-\turnover-\reprooff)$, needs $\turnover \ll 1-\reprooff$
to reduce to the bare $\turnover$ of the theorem's statement; under that
regime
$F(t) = \turnover\bigl(t - A(\reprooff)\bigr)(1+o(1))$,
which \Cref{cor:duality} places within one period of $\tau(\reprooff)$.
Replacing the discrete factors in \eqref{eq:Fexact} by $e^{-\tilde\turnover t}$
and $e^{-\alpha t}$, with rates $\tilde\turnover=\ln(1/(1-\turnover))$ and
$\alpha=\ln(1/\reprooff)$, recovers the familiar continuous-time form
$1 - [\alpha e^{-\tilde\turnover t} - \turnover e^{-\alpha t}]/(\alpha-\turnover)$;
since $1-\lambda \le \ln(1/\lambda) \le (1-\lambda)/\lambda$, the two
agree to first order in $\turnover$ and
$1-\reprooff$, and we keep the discrete form because in it the delay appears
directly as the theorem's $A(\reprooff)$, where the continuous form returns
$\tau(\reprooff)$. The consumption gap adds the immediate direct-channel step and
inherits this time path for its corpus-mediated component, since consumption is linear
in the corpus stock.

\begin{proof}[Proof of \Cref{prop:identification}]
(i) A user's consumption in arm $a$ decomposes as
$E^{a}_t = \eta_{\mathrm{e}} X^{a}_t + \eta\,\corpus_t$,
using \S\ref{sec:dynamics}'s affine consumption map,
where the exploration delivery $X^{a}_t$ differs by arm and organic
consumption depends on the corpus state $\corpus_t$, a single stock driven
by aggregate delivery and submissions. Enrollment perturbs those aggregates
by $O(p)$, but $\corpus_t$ enters both arms identically, so the perturbation
subtracts out and the contrast equals
$\eta_{\mathrm{e}}\bigl(X^{\mathrm{ctrl}}_t - X^{\mathrm{abl}}_t\bigr)
= \Delta_{\mathrm{direct}}$ for every $t$ and $N$.
(ii) Each creator population follows its own supply recursion,
$\subs^{a}_{t+1} = \subs_0 + \creatorresp(\bar V^{a})\,\subs^{a}_t$, whose
eigenvalue is that arm's loop gain $\repro^{a} = \creatorresp(\bar V^{a})$ and
where $\bar V^{a}$ is views per item under that arm's budgets; aggregate viewer
attention moves the two $\bar V^{a}$ together by $O(p)$. The per-creator
posting contrast at equilibrium is therefore
$\subs_0\bigl[(1-\repro)^{-1} - (1-\repro_{\mathrm{f}})^{-1}\bigr] + O(p)$.
The consumption generated by the induced corpus accrues to viewers of both
arms through the shared $\corpus_t$ and cancels, as in (i). The $O(p)$ term
has a determinate sign: the above-floor budget freed by flooring returns to
the shared capacity pool and redistributes over the remaining $1-p$ share, so
per-item delivery in the production arm rises relative to full rollout by a
factor of order $p/(1-p)$ times the reallocated budget share. Since
$\creatorresp$ is increasing, this inflates the measured contrast, and
$\epsilon(p) = O\bigl(p/(1-p)\bigr)$ with $\epsilon(p) \ge 0$.
(iii) Within an isolated submarket the full two-component system runs closed, so
the corpus stock expresses in the cell contrast on its own timescale
(\Cref{thm:horizon}). The final claim follows because the contrasts in (i)
and (ii) contain no corpus term at any $t$ or $N$.
\end{proof}

\subsection{Information Limits for Valuing the Corpus Multiplier}
\label{app:floors}

This section proves the information limits stated in \Cref{sec:measurement} (\Cref{thm:floor} and \Cref{cor:multiplier-rate}) and records the noise model behind them. In the co-diverted application of \S\ref{sec:probe}, the level is the direct channel and the asymptote the corpus term, so we write the gap series as
\[
y_t \;=\; \Delta_{\mathrm{direct}} + \Delta_{\mathrm{corpus}}
\bigl(1-e^{-\turnover t}\bigr) + \eta_t,
\qquad t = 1, \dots, T,
\]
with $N$ users per arm. The estimand is $\Delta_{\mathrm{corpus}}$;
$\Delta_{\mathrm{direct}}$ and the turnover $\turnover$ are unknown
nuisance parameters.

\ifextendedbuild
\begin{proof}[Proof of \Cref{lem:panel}]
\else
\begin{lemma}[Panel reduction]
\label{lem:panel}
Let per-user outcomes carry persistent user effects of arbitrary variance
$\sigma_\alpha^2$ and stationary transient noise, so that
$\mathrm{Cov}(\eta_s, \eta_t) = \tfrac{2}{N}\bigl[\sigma_\alpha^2 +
\gamma(s-t)\bigr]$. Then:
(i) because $\mathbf{1}$ is an eigenvector of the persistent component,
profiling any nuisance set containing the intercept
$\Delta_{\mathrm{direct}}$ renders every information quantity below
\emph{independent of $\sigma_\alpha^2$}, and persistent effects can only
enlarge the impossibility region of \Cref{thm:floor};
(ii) for AR(1) transient noise with parameter $\rho$ and marginal variance
$\sigma^2$, all statements hold with
$\sigma_{\mathrm{eff}}^2 = \sigma^2(1+\rho)/(1-\rho)$, up to a factor
$1 + O\bigl(\rho/((1-\rho)T)\bigr)$, by the exact tridiagonal form of the
AR(1) inverse;
(iii) stationarity is necessary: under random-walk user drift the $T^5$
information accumulation below degrades to $T^3$, because a unit root
mimics the ramp. Stationarity is diagnosable in logs: the variance of
user-mean-centered outcomes must not grow with the window.
\end{lemma}

\begin{proof}
\fi
Write the transient covariance of the arm-mean series as
$\Sigma_0 = \tfrac{2}{N}\Gamma$ with $\Gamma_{st} = \gamma(s-t)$, and the
full covariance as $\Sigma_a = \Sigma_0 + aJ$, where
$J = \mathbf{1}\mathbf{1}^{\top}$ and $a = \tfrac{2}{N}\sigma_\alpha^2$.
(i) With the intercept free, every likelihood ratio and information
quantity below depends on the data law only through the profiled precision
of its covariance, $P_\Sigma = \Sigma^{-1} - \Sigma^{-1}\mathbf{1}
(\mathbf{1}^{\top}\Sigma^{-1}\mathbf{1})^{-1}\mathbf{1}^{\top}\Sigma^{-1}$
for generic $\Sigma$.
Let $s = \mathbf{1}^{\top}\Sigma_0^{-1}\mathbf{1} > 0$. Sherman--Morrison
gives
$\Sigma_a^{-1} = \Sigma_0^{-1} - \tfrac{a}{1+as}\,\Sigma_0^{-1}J\Sigma_0^{-1}$,
hence $\Sigma_a^{-1}\mathbf{1} = \Sigma_0^{-1}\mathbf{1}/(1+as)$ and
$\mathbf{1}^{\top}\Sigma_a^{-1}\mathbf{1} = s/(1+as)$. Substituting,
\[
P_{\Sigma_a}
= \Sigma_0^{-1} - \Bigl[\tfrac{a}{1+as} + \tfrac{1}{s(1+as)}\Bigr]
  \Sigma_0^{-1}J\Sigma_0^{-1}
= \Sigma_0^{-1} - \tfrac{1}{s}\,\Sigma_0^{-1}J\Sigma_0^{-1}
= P_{\Sigma_0},
\]
identically in $a$: the profiled model is the same for every
$\sigma_\alpha^2$. For unprofiled quantities,
$\Sigma_a \succeq \Sigma_0$ in the Loewner order, so
$\Sigma_a^{-1} \preceq \Sigma_0^{-1}$ and information can only shrink,
enlarging the impossibility region.
(ii) For AR(1), $\Gamma = \sigma^2[\rho^{|s-t|}]$ admits the exact
decomposition
\[
\begin{aligned}
m^{\top}\Gamma^{-1}m
={}& \frac{(1-\rho)^2\sum_{t=1}^{T} m_t^2
  + \rho\sum_{t=2}^{T}(m_t - m_{t-1})^2}
       {\sigma^2(1-\rho^2)} \\
&+ \frac{\rho(1-\rho)\bigl(m_1^2 + m_T^2\bigr)}
       {\sigma^2(1-\rho^2)} .
\end{aligned}
\]
verified by collecting coefficients against the tridiagonal inverse. The
first term is $\sigma_{\mathrm{eff}}^{-2}\sum_t m_t^2$ with
$\sigma_{\mathrm{eff}}^2 = \sigma^2(1+\rho)/(1-\rho)$, and both
corrections vanish at $\rho = 0$. For the polynomial regressors underlying
$S_k$, where $|m_t - m_{t-1}| \le C\,|m_t|/t$ and
$m_T^2 = O\bigl(\sum_t m_t^2 / T\bigr)$, the increment term is a relative
$O\bigl(\rho/((1-\rho)^2 T^2)\bigr)$ and the boundary term a relative
$O\bigl(\rho/((1-\rho)T)\bigr)$. Every statement therefore holds with
$\sigma^2$ replaced by $\sigma_{\mathrm{eff}}^2$ up to a relative
$O\bigl(\rho/((1-\rho)T)\bigr)$.
(iii) Under random-walk drift, $\Gamma_{st} = \sigma_w^2 \min(s,t)$, and
first differencing whitens the noise while reducing the curve to its
increments. The ramp's information then accumulates from increments of the
quadratic onset, $\sum_{t\le T} t^2 \asymp T^3$, in place of
$S_4 \asymp T^5$; equivalently, level-plus-random-walk sample paths span
slowly varying trajectories that mimic the ramp, leaving identification to
the increments alone. The diagnostic in the statement follows since under
stationarity the variance of user-mean-centered outcomes is bounded in $T$,
while under a unit root it grows linearly.
\end{proof}

\noindent In what follows $\eta_t$ is accordingly taken i.i.d.\
$N(0, s^2)$ with $s^2 = 2\sigma_{\mathrm{eff}}^2/N$, and
$S_k \triangleq \sum_{t \le T} t^k$.

\begin{proof}[Proof of \Cref{thm:floor}]
Let the true parameters be $\Delta_{\mathrm{direct}}$, $\Delta_0$ (the
statement's true $\Delta_\infty$) and
$\turnover_0$, and write $1 - e^{-x} = x\,\phi(x)$ with $\phi(x) = (1-e^{-x})/x$,
$\phi(0) = 1$. Then $0 \le -\phi'(x) \le \tfrac12$ for all $x \ge 0$:
$-x^2\phi'(x) = 1-(1+x)e^{-x}$, whose derivative is $x e^{-x} \in [0, x]$,
so $0 \le 1-(1+x)e^{-x} \le x^2/2$. For $M \in (\Delta_0, \infty]$ consider
the \emph{matched-product} alternative
$(\Delta_{\mathrm{direct}}, M, \turnover_M)$ with
$\turnover_M = \turnover_0 \Delta_0 / M$; the case $M = \infty$ is the
linear curve $\Delta_{\mathrm{direct}} + \Delta_0\turnover_0\,t$. With
$\beta = \Delta_0 \turnover_0$, the mean curves differ by
\[
\begin{aligned}
|f_M(t)-f_0(t)|
&= \beta t\,\bigl|\phi(\turnover_M t)-\phi(\turnover_0 t)\bigr| \\
&\le \tfrac12 \beta t^2(\turnover_0-\turnover_M)
 \le \tfrac12 \Delta_0\turnover_0^2t^2 ,
\end{aligned}
\]
uniformly in $M$. Hence $D_M^2 = \sum_t (f_M - f_0)^2 \le
\tfrac14 \Delta_0^2 \turnover_0^4 S_4 = \mathcal{C}^2 s^2 \le s^2$, so for
the product Gaussian measures
$\mathrm{KL}(P_0 \| P_M) = D_M^2/(2s^2) \le \tfrac12$ and
$\mathrm{TV} \le \sqrt{\mathrm{KL}/2} \le \tfrac12$, simultaneously for
every $M$. Coverage at $\theta_M$ transfers:
$P_0(M \in \mathrm{CI}) \ge P_M(M \in \mathrm{CI}) - \mathrm{TV} \ge
\tfrac12 - \alpha$. The events $\{U \ge M\} \supseteq \{M \in \mathrm{CI}\}$
decrease to $\{U = +\infty\}$ as $M \uparrow \infty$, and monotone
convergence gives the claim.
\end{proof}

\noindent The certified constant is conservative in two removable ways.
Replacing Pinsker's inequality with the exact Gaussian total variation,
$\mathrm{TV}(P_0, P_M) = 2\Phi\bigl(D_M/(2s)\bigr) - 1$ with $\Phi$ the
standard normal c.d.f., certifies the same conclusion up to
$\mathcal{C} \le 2\Phi^{-1}(3/4) \approx 1.35$; and allowing the
alternative to re-profile the level and slope shrinks $D_M$ further, since
the leading difference $\propto t^2$ is far from orthogonal to
$\mathrm{span}\{1, t\}$. Both enlargements extend the impossibility region,
so the theorem's direction is unaffected; the calibrated Monte Carlo of
\Cref{fig:floor-demo} accordingly places the practical estimability
transition roughly an order of magnitude above the certified threshold.

\begin{proof}[Proof of \Cref{cor:multiplier-rate}]
(i), lower bound: choose $M$ in the proof of \Cref{thm:floor} so that $D_M=s$.
The pair is then separated by
\[
\begin{aligned}
|M-\Delta_0| &\gtrsim \frac{s}{\turnover^2\sqrt{S_4}}, \\
|M-\Delta_0| &\asymp \frac{s}{\turnover^2T^{5/2}}.
\end{aligned}
\]
while $\mathrm{TV} \le \tfrac12$, and the
standard two-point reduction converts indistinguishability into the risk
bound.
For (i)'s attainment claim: expand the mean curve's derivatives to second
order in $\turnover t$ and profile the intercept and $\turnover$. The Gaussian
Fisher information then leaves
\[
 I^*(\Delta_{\mathrm{corpus}})
 = \frac{\turnover^4 T^5}{720s^2}\bigl(1+O(\turnover T)\bigr).
\]
The Gaussian regression has an analytic mean and nonsingular information at
every interior point with $\turnover>0$, so standard parametric theory gives
asymptotic attainment by profile maximum likelihood. Inverting $I^*$ for a
target relative error gives the $N^{-1/5}$ law.
(ii) Delta method on $\Delta_{\mathrm{corpus}} = \beta/\turnover$, with
$\hat\beta$ the intercept-profiled slope: the slope's information is
$\turnover^2 T^3/(12 s^2)$ to leading order, so the error ratio between
(ii) and the known-$\turnover$ problem is
$\sqrt{720/12}/(\turnover T) \approx 7.7/(\turnover T)$, which the external
estimate removes.
\end{proof}

\begin{remark}[Estimator choice]
A numerical check at $\turnover T \le 0.3$ confirms that profile likelihood
attains the bound at the probe's scale. The quadratic plug-in estimator
$\hat\Delta = \hat u^2/(2\hat v)$ from a polynomial fit is truncation-biased
at fixed $\turnover T$. Wherever the single-rate form is fit, profile likelihood
is used for this reason. The probe's primary
analysis avoids the single-rate form entirely, regressing on the measured
age--yield kernel and the posting-path convolution (\S\ref{sec:probe}).
\end{remark}

\section{Omitted Results and the Co-Diverted Estimator}
\label{app:extended}

This appendix states the results omitted from the main text for length and
details the estimator used in the co-diverted analysis. Proofs are in
\Cref{app:props}. It is self-contained given the notation of
\S\ref{sec:budget-problem}.

\subsection{Finite-horizon information limits}

The following two results are stated without proof in
\S\ref{sec:measurement}.

Detecting a nonzero finite-horizon effect is easier than estimating its
asymptote. Let $y_t$ be the observed between-arm contrast in a corpus-stock
metric at horizon $t$. Suppose it has an unknown level plus a single slow
component,
\[
y_t \;=\; c + \Delta_\infty\bigl(1 - e^{-\turnover t}\bigr) + \varepsilon_t,
\qquad t = 1, \dots, T,
\]
with $\varepsilon_t$ i.i.d.\ $N(0, s^2)$, $s^2 = 2\sigma_{\mathrm{eff}}^2/N$ for
$N$ users per arm, where $\sigma_{\mathrm{eff}}$ is the effective per-user
transient-noise scale (\Cref{lem:panel}, Appendix~\ref{app:floors}, which also
shows that persistent user heterogeneity costs nothing once the level is
profiled), and $S_4 = \sum_{t \le T} t^4$. Both the level $c$ and the
rate $\turnover$ are unknown nuisance parameters.

\begin{theorem}[No finite upper confidence bound below the curvature
threshold]
\label{thm:floor}
Define the curvature signal-to-noise ratio
$\mathcal{C} = \Delta_\infty\,\turnover^2\sqrt{S_4}/(2s)$.
If $\mathcal{C} \le 1$, then any confidence procedure for $\Delta_\infty$
with uniform coverage $1-\alpha$ over positive amplitudes and rates
satisfies, under the true parameters,
\[
  P(\text{upper endpoint} = +\infty) \;\ge\; \tfrac12 - \alpha .
\]
\end{theorem}

\begin{corollary}[Horizon--sample law for the asymptote]
\label{cor:multiplier-rate}
(i) Any estimator of $\Delta_\infty$ errs by
$\gtrsim s/(\turnover^2 T^{5/2})$ on some parameter within total variation
$\tfrac12$ of the truth. To leading order in $\turnover T$ this rate is
attained by profile likelihood, so a fixed relative-error target requires
$T \propto N^{-1/5}$: a tenfold increase in traffic shortens the required
horizon by a factor of $1.6$.
(ii) With $\turnover$ estimated from auxiliary data to relative precision
$\rho_{\turnover}$, $\Delta_\infty = \beta/\turnover$ inherits the slope's law:
relative error $\approx \sqrt{\smash[b]{\mathrm{relerr}(\hat\beta)^2 +
\rho_{\turnover}^2}}$, the horizon law improves to $T \propto N^{-1/3}$, and an error
inflation of order $1/(\turnover T)$ is removed.
\end{corollary}
\noindent Proofs are in Appendix~\ref{app:floors}. The theorem uses an
elementary two-point argument along the family of curves with matched
product $\Delta_\infty\turnover$, which includes the straight line (the
asymptote-$\infty$ limit) within total variation $\tfrac12$ of the
truth whenever $\mathcal{C} \le 1$. For the ecosystem, the slow component is
the corpus stock: the regimes in which exploration is most valuable
(\Cref{cor:duality}) are those in which its asymptotic value can, at
feasible horizons, admit only an increment lower bound. \S\ref{sec:probe} runs
exactly this estimation problem with a horizon-matched endpoint.

\begin{figure}[t]
  \centering
  \includegraphics[width=\linewidth]{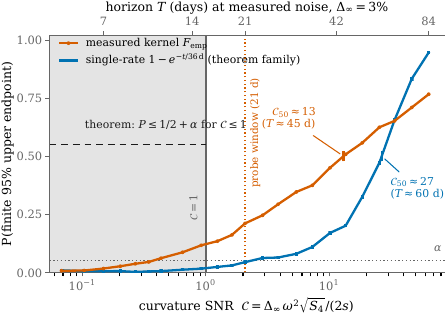}
  \caption{Finite-upper-endpoint probability for a $95\%$ profile procedure
  in semi-synthetic Monte Carlo at the measured noise scale. Blue is the
  exact single-rate family of \Cref{thm:floor}. Its $\mathcal C\le1$ region
  obeys the theorem's $\tfrac12+\alpha$ cap. Orange uses the measured
  age--yield kernel under free timescale dilation. It shares the horizontal
  coordinate through the kernel's 36-day single-rate equivalent. At the
  probe's 21-day window ($\mathcal C\approx2$), its finite-endpoint
  probability is $0.21$. Markers give each curve's median crossing.}
  \label{fig:floor-demo}
\end{figure}

\subsection{The estimator used in the co-diverted analysis}
\label{app:estimator}

The results above analyze the design-stage problem, where the corpus timescale
is unknown. The estimator used in \S\ref{sec:probe} instead pins the response
shape from historical platform data. For horizon $t$ in the co-diverted
contrast, the model is
\[
\begin{aligned}
  y_t ={}& c + \delta_t + \Delta_{\mathrm{direct}}\,\mathbf{1}\{t \ge 0\} \\
  &+ \Delta_{\mathrm{corpus}}\,(k_{\mathrm{emp}} \ast \pi)(t) + \varepsilon_t,
\end{aligned}
\]
where $k_{\mathrm{emp}}$ is the measured age--yield kernel, $\pi$ the measured
in-cell posting path normalized to unit long-run level, $\delta_t$ the drift
imported from the concurrent one-sided ablation (\S\ref{sec:probe}), and
$\varepsilon_t$ independent with the per-estimate sampling variances.
Writing $F_{\mathrm{emp}}$ for the kernel's normalized cumulative response,
the eleven monthly cross-sections give
\[
F_{\mathrm{emp}}(7)=0.471,\quad F_{\mathrm{emp}}(21)=0.583,\quad
F_{\mathrm{emp}}(90)=0.755.
\]
These values display the front-loaded, non-exponential shape behind the
36-day single-rate equivalent used in the main text.

Interpreting $k_{\mathrm{emp}}$ as the intervention response uses four
assumptions. (A1) A change in exploration delivery scales cohort inflow without changing the
normalized age--yield shape. (A2) Ranker competition between cohorts leaves
that shape unchanged at the probe's perturbation size. (A3) The historical
cross-sections transport to the cells. Both run the production ranker, and the
dispersion across the eleven cross-sections enters the variance below. (A4)
The observed in-cell posting path is the relevant intervention input.
Once the ramp is estimable, (A1)--(A2) can be assessed by comparing the
cell's own age profile with $k_{\mathrm{emp}}$.

With $k_{\mathrm{emp}}$ and $\pi$ pinned and $\delta_t$ imported, the model is
linear in the intercept $c$ and the two coefficients
$\Delta_{\mathrm{direct}}$ and $\Delta_{\mathrm{corpus}}$. The estimator is
weighted least squares under the sampling variances. The step and
ramp regressors are linearly independent once three or more post-onset
estimates exist, which gives algebraic identification. Reporting also requires
the stability and interval checks described below. Uncertainty combines
three components, summed as variances under independence across sources.
(i) Sampling error uses the per-estimate variances, with estimates over
non-overlapping windows treated as independent. Shared-user serial correlation
is largely removed by the pre-period differencing, and as a check the
variances are inflated by the estimated lag-one autocorrelation of the
residual series. (ii) Kernel uncertainty treats the eleven cross-sectional
kernels as draws: the fit is repeated per kernel and the across-kernel
variance of $\hat\Delta_{\mathrm{corpus}}$ is added. (iii) Drift enters as a
Gaussian prior centered on the imported series with its reported margins,
propagated through the weighted fit. Coverage of the combined interval is
checked on synthetic data with known truth under the specified procedure. In
300 replicates of each nondegenerate scenario, combined-gap coverage is
$93.7\%$ (Monte Carlo s.e. $1.4$ points), while the lower-bound construction
is valid in $96$--$97\%$. If the posting path is exactly zero, the supply
regressor is dropped and coverage falls to $89.0\%$. The reported analysis
uses the nonzero-path specification, whose kernel--supply regressor correlation
is $0.994$. This nominally passes the pre-specified collinearity cutoff of
$0.995$, fixed before estimation.
Moreover, the primary interval includes zero and large offsetting step and ramp
coefficients expose intercept--kernel near-collinearity. The pre-specified joint rule
rejects a two-sided ceiling estimate. The main text reports the
well-conditioned robustness intervals instead. The estimation procedure
distinguishes two targets. Conditional on (A1)--(A4), the WLS coefficient is a
two-sided estimate of $\Delta_{\mathrm{corpus}}$. The shape-robust companion
uses an increment. An envelope regressor would leave the coefficient unbounded
once the direct-step nuisance is profiled. For horizons
$t_1 < t_2$ after the exploration window, the drift-adjusted increment has mean
$\Delta_{\mathrm{corpus}}\,[F(t_2)-F(t_1)]$: the direct step and the level
cancel from the difference, and under the stock model $F$ is nondecreasing
with $F \le 1$, so $0 \le F(t_2)-F(t_1) \le 1$ for \emph{any} admissible
response shape. The increment has the same sign whenever the stock
changes and never exceeds $\lvert\Delta_{\mathrm{corpus}}\rvert$. The
nonnegative-stock model imposes $\Delta_{\mathrm{corpus}} \ge 0$ on the
gross-flow target (\Cref{prop:gradient}). This direction was fixed before the
ramp was estimable, and we use it only for the model-restricted lower bound.
The policy-relevant net asymptote in the main text is reported without this
restriction and remains unsigned. Under the directional restriction, the
robust bound is the lower endpoint of the one-sided $95\%$
interval on the drift-adjusted increment. If the
estimate contradicts that direction, we report the two-sided $95\%$
interval and claim a nonzero corpus term only if it excludes zero. The bound
costs sharpness, paying $1-[F(t_2)-F(t_1)]$ of the asymptote, but assumes
nothing about the shape beyond monotonicity.

\section{Additional Empirical Analyses}
\label{app:timepath}

\subsection{The Observed Viewer Time Path}

\Cref{sec:viewer-ablation} rests part of its identification argument on the
shape of the viewer ablation over time. The main text summarizes that path in
\Cref{fig:observed-evidence}. \Cref{fig:period} shows seven cumulative estimates in faint lines and their
six non-overlapping period differences in solid lines. The final cumulative
estimates reproduce the video-view and view-time entries in
Table~\ref{tab:ablations}, Panel A.

Three features matter, and one of them is easy to misread from the
cumulative series alone.

\emph{No sign flip or ramp.} Each series holds its sign across all
seven estimates, and neither exhibits a sustained upward ramp. A growing
corpus contribution would build monotonically on the turnover timescale
(\Cref{thm:horizon}). Nothing here does. This is the feature
\Cref{sec:viewer-ablation} uses to read the gap as the direct, per-user
effect.

\emph{Cumulative smoothing hides attenuation.}
Video views rise from $+2.02$ to a peak of $+2.32$ around month seven, then
fall to $+1.74$. Read this way the
excursion looks small, $0.58$ percentage points against effects of $1.7$ to
$2.3$. A cumulative reading necessarily looks small because each estimate
contains every earlier one. Consecutive points are
constrained to agree whatever the underlying effect is doing, making a
cumulative series the wrong instrument for a claim about
shape. \Cref{fig:period} differences the same seven estimates into six
non-overlapping periods, which share no data.
\begin{equation*}
  \Delta_{(t_1,t_2]} \;=\; \frac{c_2 t_2 - c_1 t_1}{t_2 - t_1},
\end{equation*}
exact when the cumulative denominator grows in proportion to elapsed time,
i.e., per-period exposure is stable. With balanced arms of fixed membership
and equal-length windows the deviation is second order. Without covariances
between nested cumulative estimates, the plotted bands scale
the cumulative margins under independent per-period noise, which overstates
increment variance when period-level variances are stable. We label
the bands approximate. On that reading video views
hold between $+2.27$ and $+2.45$ for the first three periods and then fall
to $+0.13$, and view time deepens to $-2.87$ before
attenuating to $-1.14$. The movement is several times larger than
the cumulative series suggests.

Both shapes lie outside the corpus contribution, which a one-sided design cannot
express at any horizon (\Cref{prop:identification}(i)), so the reading of
\S\ref{sec:viewer-ablation} is unaffected. The path is a stable regime followed
by attenuation, which coincides with a decline in
exploration's share of feed consumption over the same span. An ablation's
effect size scales with the share of delivery it removes, and that share is
an operating choice. These figures report an operating-point price.

\begin{figure}[t]
  \centering
  \includegraphics[width=\linewidth]{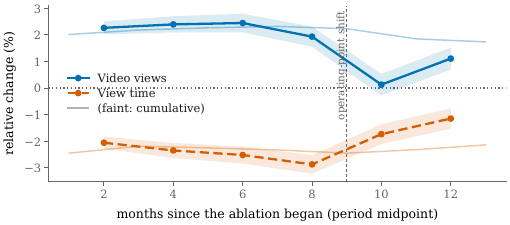}
  \caption{The same seven cumulative estimates differenced into six non-overlapping
  periods (solid, with approximate bands), against the cumulative reading
  (faint). The dashed vertical marker is the operating-point shift both
  experiments turn at (\S\ref{sec:viewer-ablation}). Positive $=$ higher under production. Consecutive periods share
  no data, so their shape is the effect's shape. The cumulative series is
  smooth by construction. Period bands are approximate upper bounds derived
  from the cumulative margins under independent per-period noise, scaling
  the cumulative margin by $\sqrt{t_2/(t_2-t_1)}$. No closed-form margin
  for nested increments is available from the per-horizon estimates.}
  \label{fig:period}
\end{figure}

\emph{Precision without convergence.} The confidence
bands narrow monotonically as the cumulative sample grows, from $\pm0.23$
to $\pm0.16$ percentage points on video views. The point estimate moves by
more than that narrowing over the same window, which is
\Cref{prop:detection}(i) in miniature. More data measures the system's
state at the time of measurement more precisely without pinning down a
quantity that is itself drifting.

\Cref{fig:creator-period} repeats the period reading for the concurrent
creator ablation's \emph{consumption} series, the second, independently
randomized witness of the operating-point shift cited in
\S\ref{sec:viewer-ablation}. Period effects on video views hold near
$+1\%$ up to the same boundary ($+1.11\%$ in the final period before it)
and turn negative after.

\begin{figure}[t]
  \centering
  \includegraphics[width=\linewidth]{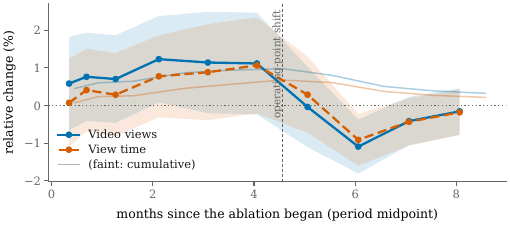}
  \caption{The creator ablation's consumption series read per period
  (solid, approximate bands) against its cumulative reading (faint), as in
  \Cref{fig:period}. The dashed marker is the same operating-point boundary
  both experiments turn at. Period effects on video views are positive up to it
  and negative after, the coincident break cited in
  \S\ref{sec:viewer-ablation}.}
  \label{fig:creator-period}
\end{figure}

\subsection{Budget-Matched Reallocation}
\label{app:matched-budget}

A separate randomized co-diverted experiment tests reallocation
(\S\ref{sec:reallocation}). As in \S\ref{sec:probe}, viewers and creators
were assigned to matched isolated cells. Intervals condition on these realized
cells. The design holds the nominal total
budget fixed and spreads exploration across a broader set of videos. Promotion
thresholds implement the shift, and a spend comparison on the
penultimate day finds realized budget per creator matched within
approximately $3\%$, the reallocated arm slightly lower, so any residual
dose imbalance biases its posting response downward. Snapshot video
counts differ because reallocation changes pool residence. This one-day
diagnostic bounds the observed residual.
Window-wide equivalence remains unverified, so the estimate is the randomized
arm contrast, leaving pure breadth elasticity open. The experiment
runs for two weeks, a horizon matched to
the fast supply margin it targets. The production-versus-reallocation
contrast and the two-week horizon belong to the original experiment design,
which was run independently of this study. A second planned reallocation arm
is omitted because a configuration audit showed that it did not implement the
intended budget-matched contrast. The intervention configuration determined
this exclusion before the outcomes were considered. Over the full window, creators
posting at least once increase by $0.77\%\pm0.42$, and videos posted per creator
increase by $0.92\%\pm0.87$. Viewer-side consumption shows no detectable change
(view time $+0.01\%\pm0.15$, video views $-0.06\%\pm0.15$, and video favorites
$+0.15\%\pm0.59$). The two-week window captures only the
fast margin. By \Cref{thm:horizon}, the corpus consequence of the induced
supply is attenuated and its asymptote unidentified in this window. The result
identifies the direction of the short-horizon trade-off. Long-run value and
the allocation rule remain open.

\subsection{Feasible-Horizon Corpus Flow in the Co-Diverted Cells}
\label{app:cohortledger}

Splitting each co-diverted cell's consumption by video posting cohort and
video age at viewing isolates organic consumption of videos posted during the
experiment after their exploration windows. By construction, exploration can
no longer deliver these videos. Among videos posted by creators assigned to
the same cell as the viewers, production raises organic view time by $37.9\%$
$[37.1, 38.7]$ relative to the floored cell and raises video views by $41.7\%$.
Videos posted before the switch show no effect once the exploration-window
period is excluded. All subsequent estimates for pre-experiment videos are
within $\pm 3\%$. The pooled reading that includes the exploration-window period
falls outside that band. We report both versions. \Cref{fig:incell-supp}
plots the daily series against the placebo band. Video favorites per viewer on
in-cell videos rise $29.8\%$ $[24.3, 35.3]$. Per view, marginal corpus videos
receive $\approx\!0.91\times$ the incumbent viewing duration and have
$\approx\!0.71\times$ the incumbent favorite rate. Both are marginal ratios
of the outcome change to the change in video views, relative to the incumbent
per-view rate. The
affected slice consists of in-cell videos posted during the experiment and
viewed after their exploration windows. It carries $0.3$--$0.5\%$ of a cell's
consumption at this horizon. Together, all in-cell videos account for
$8$--$14\%$ of a cell's consumption. The $+37.9\%$ effect on
it moves the aggregate contrast by roughly $0.1$--$0.2$ percentage points,
a magnitude consistent with the whole-cell null.

\emph{Inferential status.} The decomposition was discovered in an exploratory
analysis of a $1\%$ user sample. Applying the same endpoints, estimators, and
windows to the full population reproduces the numerical pattern without
creating an independent confirmatory test. The pooled pre-experiment cohort
also misses the placebo band when the exploration-window period is included.
Excluding that period, during which the intervention can still affect delivery,
restores the placebo pattern. We report both versions and retain exploratory
status. A shape-based step-plus-ramp decomposition of the net term is also
underidentified because its collinearity criterion fails at the available
horizon, so we report a bound. Accordingly, the
main text uses this decomposition to measure feasible-horizon gross corpus
flow. The sign of aggregate corpus value remains unresolved.

\begin{figure}[t]
  \centering
  \includegraphics[width=\linewidth]{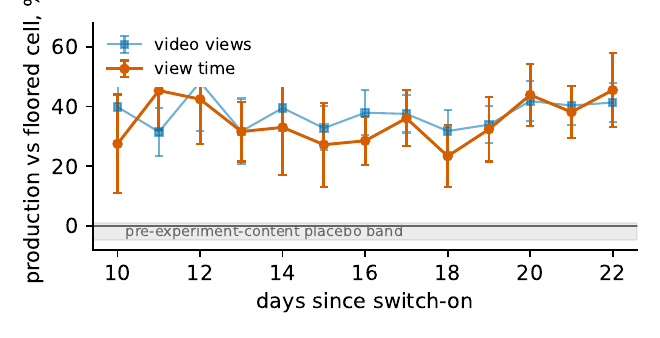}
  \caption{In-cell organic video views and view time for videos posted during
  the experiment and viewed after their exploration windows, production versus
  floored cell, daily with $95\%$ intervals (user-level sample). The pooled
  full-population view-time contrast is $+37.9\%$ $[37.1, 38.7]$. The gray band is the placebo interval from
  videos posted before the experiment.}
  \label{fig:incell-supp}
\end{figure}

\section{Worked Illustrations}
\label{app:illustrations}

The paper states its results as bounds and scaling laws. This appendix works
four of them through numerically at one illustrative parameter setting and
gives each an operational reading. Every parameter is chosen for legibility
and differs from production values. The parameters stated below determine
every figure in this section.

\paragraph{The running example.}
One period is a submission-feedback cycle. We take corpus turnover
$\turnover = 0.005$ per period, so the corpus half-life is
$\ln 2/\ln(1/(1-\turnover)) \approx 138$ periods and
$\tau_{\turnover} = 1/\turnover = 200$. We use a post-switch loop gain
$\reprooff = 0.9$, so one supply-loop multiplier is
$A(\reprooff) = 1/(1-\reprooff) = 10$ periods. The two timescales are separated
by a factor of $20$, the regime \S\ref{sec:dynamics} assumes.

\begin{figure}[t]
  \centering
  \includegraphics[width=\linewidth]{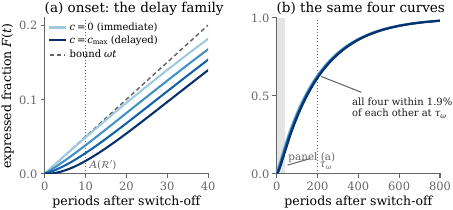}
  \caption{\Cref{thm:horizon}'s delay family, from \eqref{eq:Fexact}
  (\S\ref{app:props}) at
  $\turnover=0.005$, $\reprooff=0.9$. Four values of the interpolating
  coefficient, $c/c_{\max} \in \{0, \tfrac13, \tfrac23, 1\}$, light to dark.
  \textbf{(a)} The onset differs. At $c=0$ the corpus responds at rate
  $\turnover$ from the first period. At $c=c_{\max}$ the first period
  expresses nothing, and the ramp extrapolates back to zero at exactly
  $A(\reprooff)$. \textbf{(b)} The same four curves over the corpus timescale,
  where the distinction has washed out.}
  \label{fig:supp-two-ends}
\end{figure}

\subsection{Theorem 1: the cost of delay}

\Cref{thm:horizon} leaves one coefficient free, $c \in
[0, \turnover/(1-\turnover-\reprooff)]$. How much turns on it? At the running
parameters $c_{\max} = 0.0526$, and the answer splits by horizon.

\Cref{fig:supp-two-ends}(a) shows the first $40$ periods. The two ends have
genuinely different shapes. At $c=0$, $F(1) = \turnover$ exactly and the
response is linear from the start. At $c=c_{\max}$, $F(1) = 0$ exactly, the
onset is quadratic, and the ramp extrapolates back to zero at
$t_0 = 10.000$ periods, which is $A(\reprooff)$ to four decimals, as the
algebra behind \eqref{eq:Fexact} in \S\ref{app:props} requires. That
intercept is what a fitted ramp reports. The exact $F(10) = 0.017$, essentially
the fast term the ramp discards. A three-week window would report materially
different numbers depending on where a platform sits.

\Cref{fig:supp-two-ends}(b) shows the same four curves out to $800$ periods.
By $\tau_{\turnover}$ they agree to within $1.9\%$ of the asymptote, and by
$t=800$ to within $0.1\%$. The delay is a transient on the fast timescale.
The approach to the asymptote is governed by $\turnover$ alone.

\paragraph{Planning consequence.}
Two consequences point in opposite directions. Planning requires no estimate
of $c$. The horizon is set by $\turnover$, estimable from turnover data without running
the experiment. Reading a short experiment does require it, since the
attenuation factor depends on $c$. A short window's estimate is attenuated by an amount
the model leaves unspecified, which is why \S\ref{sec:probe} fits the delayed
onset from data.

\begin{figure}[t]
  \centering
  \includegraphics[width=\linewidth]{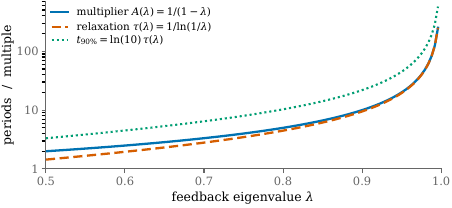}
  \caption{\Cref{cor:duality}: the multiplier a feedback loop delivers and its
  relaxation time coincide to within one period. Both are plotted against a
  single axis. A target expression horizon adds the factor set by that target.
  In particular, $t_{90\%} = \ln(10)\,\tau(\lambda)$.}
  \label{fig:supp-duality}
\end{figure}

\subsection{Corollary 1: reading value off the horizon axis}

\Cref{cor:duality} says $A(\lambda)$ and $\tau(\lambda)$ coincide to within
one period. \Cref{fig:supp-duality} makes the coincidence visible: the gap
$A - \tau$ is $0.56$ periods at $\lambda = 0.5$ and $0.50$ at
$\lambda = 0.99$, never exceeding one, with the ratio approaching $1$.

A loop with $\lambda = 0.9$ delivers
a $10\times$ multiplier and takes $9.5$ periods to express $63\%$ of a
switch, $21.9$ to express $90\%$. A loop with $\lambda = 0.99$ delivers
$100\times$ and takes $99.5$ and $229$ periods respectively.

\paragraph{Feasibility check.}
\Cref{fig:supp-duality} doubles as a feasibility check on a research plan. Read the
multiplier a proposal hopes to demonstrate off the vertical axis, find the
$\lambda$ that delivers it, and read off $t_{90\%}$. That is the shortest
honest study. A plan to establish a $50\times$ ecosystem multiplier inside
one turn of the corpus is internally inconsistent,
because the multiplier is achieved \emph{by} the slowness. This is also the
argument for the prescription of \S\ref{sec:intro}: since the multiplier and
the delay cannot be separated, a mechanism that must decide quickly has to
\emph{import} the slow terms. The remaining question is whether the offline
estimate is good enough.

\begin{figure}[t]
  \centering
  \includegraphics[width=\linewidth]{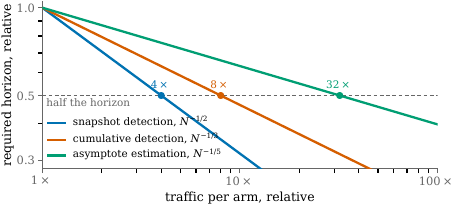}
  \caption{\Cref{prop:detection} and \Cref{cor:multiplier-rate}: required
  horizon against traffic per arm, both relative, for the three exponents
  the paper derives. Markers give the traffic multiple that halves the
  horizon.}
  \label{fig:supp-scale}
\end{figure}

\subsection{Proposition 2 and \Cref{cor:multiplier-rate}: pricing a traffic
increase}

The three exponents in the paper ($N^{-1/2}$ for snapshot detection,
$N^{-1/3}$ for cumulative detection, $N^{-1/5}$ for estimating the
asymptote) are easier to act on inverted. \Cref{fig:supp-scale} plots the
horizon a given traffic multiple buys. Tenfold traffic shortens the required
horizon by $3.2\times$, $2.2\times$, or $1.6\times$ respectively; halving
the horizon costs $4\times$, $8\times$, or $32\times$ the traffic.

\paragraph{Pricing a traffic ramp.}
This prices the standard proposal to fix a slow measurement by ramping the
experiment. Going from $1\%$ to $10\%$ of traffic is tenfold, and buys at
most $3.2\times$, but only if the estimand is mere detection of an
attenuated gap under snapshot analysis. If the estimand is the asymptotic
multiplier, the same ramp buys $1.6\times$, and reaching half the horizon
would require $32\times$ the traffic. The exponent is the binding constraint.
The remaining lever is
calendar time or an external estimate of $\turnover$, which by
\Cref{cor:multiplier-rate}(ii) moves the law from $N^{-1/5}$ to $N^{-1/3}$
and is the cheapest available improvement.

\begin{figure}[t]
  \centering
  \includegraphics[width=\linewidth]{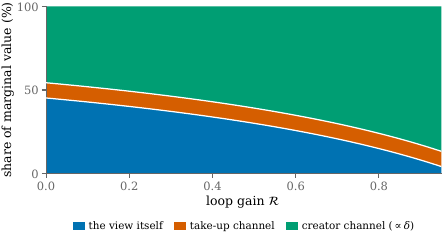}
  \caption{\Cref{prop:gradient}'s three channels as shares of the marginal
  value of a view, against loop gain $\repro$. Illustrative units
  ($\eta_{\mathrm{e}} = \eta_{\mathrm{o}} = 1$, $\Vorg = 10$,
  $\bar q = 0.1$, $\bar\budget = 100$, $q' = 0.01$, $\sensitivity = 0.01$).
  The shares depend on these. The monotonicity is structural. Stack order is
  bottom to top as listed.}
  \label{fig:supp-channels}
\end{figure}

\subsection{Proposition 1: which channel carries the value}

\Cref{prop:gradient} separates the marginal value of a view into a direct
creator channel, the view itself, and a take-up channel.
\Cref{fig:supp-channels} plots the three as shares against the loop gain.
At $\repro = 0$ the creator channel is $46\%$ of the total, rising to
$70\%$ at $\repro = 0.7$ and $87\%$ at $\repro = 0.95$. Counting the
$\sensitivity$-dependent part of the take-up channel as well, the share of
marginal value omitted from published budget objectives in this class runs from
$50\%$ to $95\%$ across the same range.

The monotonicity is structural across choices of illustrative units.
Both the take-up and creator channels carry the submission value
$v^* \propto 1/(1-\repro)$. The view-itself term lacks that multiplier, so as
$\repro \to 1$ the view-itself term's share vanishes and the creator channel
approaches $1/(1+q'\Vorg)$, which is $91\%$ here. Only that limit and the
crossover location depend on the chosen $q'\Vorg$.

\paragraph{Design consequence.}
Read against \Cref{cor:duality}, this is the paper's design argument: the
omitted channel dominates at high $\repro$, which is exactly where it is
slowest to measure. An allocator scoring only viewer-side signals discards the
majority of marginal value where that majority is largest, and the
mechanism's own feedback cannot reveal the loss (\Cref{thm:horizon}).

\end{document}